\documentclass[fleqn,usenatbib]{mnras}

\usepackage{newtxtext,newtxmath}

\usepackage[T1]{fontenc}

\DeclareRobustCommand{\VAN}[3]{#2}
\let\VANthebibliography\thebibliography
\def\thebibliography{\DeclareRobustCommand{\VAN}[3]{##3}\VANthebibliography}

\usepackage{graphicx}	
\usepackage{amsmath}	
 
\usepackage{amssymb}	

\title[Two fluid shocks]{Partially-ionised two-fluid shocks with collisional and radiative ionisation and recombination - multi-level hydrogen model}

\author[B. Snow, M. K. Druett, \& A. Hillier]{
B. Snow,$^{1}$\thanks{E-mail: b.snow@exeter.ac.uk}
M. K. Druett$^{2}$
and A. Hillier$^{1}$
\\
$^{1}$University of Exeter, Exeter, EX4 4QF, UK  \\
$^{2}$Centre for mathematical Plasma Astrophysics, KU Leuven, 3001 Leuven, Belgium
}

\date{Accepted XXX. Received YYY; in original form ZZZ}

\pubyear{2021}

\begin{document}
\label{firstpage}
\pagerange{\pageref{firstpage}--\pageref{lastpage}}
\maketitle

\begin{abstract}
   {Explosive phenomena are known to trigger a wealth of shocks in warm plasma environments, including the solar chromosphere and molecular clouds where the medium consists of both ionised and neutral species. Partial ionisation is critical in determining the behaviour of shocks, since the ions and neutrals locally decouple, allowing for substructure to exist within the shock. Accurately modelling partially ionised shocks requires careful treatment of the ionised and neutral species, and their interactions. }
   {}
   {Here we study a partially-ionised switch-off slow-mode shock using a multi-level hydrogen model with both collisional and radiative ionisation and recombination rates that are implemented into the two-fluid (P\underline{I}P) code, and study physical parameters that are typical of the solar chromosphere. 
   }
   {The multi-level hydrogen model differs significantly from MHD solutions due to the macroscopic thermal energy loss during collisional ionisation. 
   In particular, the plasma temperature both post-shock and within the finite-width is significantly cooler that the post-shock MHD temperature. Furthermore, in the mid to lower chromosphere, shocks feature far greater compression then their single-fluid MHD analogues. 
   The decreased temperature and increased compression reveal the importance of non-equilibrium ionised in the thermal evolution of shocks in partially ionised media. 
   Since partially ionised shocks are not accurately described by the Rankine-Hugoniot shock jump conditions, it may be incorrect to use these to infer properties of lower atmospheric shocks.}
\end{abstract}




\section{Introduction}

The lower solar atmosphere is rife with well-observed shocks that are driven, for example, through magnetic reconnection \citep{Petschek1964,Pontieu2007,Rouppe2007}, and wave steepening \citep{Suematsu1982,Felipe2010,Rajaguru2019}. 
For a fully ionised media, the MHD equations can be used to model shock jumps \citep{Tidman1971}. Shock jump equations can be derived to determine all possible stable shock transitions given a few properties of the upstream media \citep{Hau1989}. 
These jump conditions have been used to classify shocks that are observed in the upper solar atmosphere \citep[e.g.,][]{Bemporad2010}. 
However, single-fluid ideal MHD modelling becomes far less applicable to shocks in the lower solar atmosphere, where the atmospheric conditions result in a partially-ionised system, consisting of both ionised and neutral species.

In idealised MHD models, a shock exists as a highly compressible feature with a discontinuous jump in plasma properties and features an increase in temperature and density across the shock. However, shocks in a partially-ionised system can behave very differently than their MHD counterparts. In the two-fluid partially-ionised model, the ionised and neutral species are coupled either side of the shock, however the species decouple and recouple within the shock, leading to a finite shock width determined by the level of coupling in the system \citep{Hillier2016}. In stratified media, the finite shock width can become comparable to the  pressure scale-height of the system \citep{Snow2020}. 
Within the finite shock width, substructure can become important, leading to the formation of additional shocks within the larger-scale shock, localised heating and potential observational measures \citep[see reviews of][]{Ballester2018,Hillier2022}.

At the shock interface, the local ionisation rates exceed background rates by several orders of magnitude \citep{Carlsson2002}. As such, ionisation and recombination are critical in determining the variations in the ionisation fraction and hence the finite width of shocks and the behaviours inside it. 
During collisional ionisation, the free electron expends its kinetic energy to release the bound electron, and hence there is a macroscopic thermal energy loss from the charged species that acts as a non-adiabatic cooling term. 
Postshock (and inside the finite width of the shock), the temperature and density increase, driving up the ionisation rates and therefore increasing the collisional ionisation losses. 

A relatively simple way of modelling collisional ionisation and recombination is to use empirically determined rates \citep[e.g.,][]{Voronov1997,Smirnov2003}. Following this model, the ionisation energy lost across the shock is calculable and this system was studied extensively in \cite{Snow2021}.
A consequence of this model is that the fluid energy equation is no longer conserved (since energy can enter or leave the system through radiative losses or heating) however a semi-analytical determination of shock jumps can be derived for shocks that are modelled using radiative losses. 
It was shown analytically that a compressible shock must cool across the interface in the model of \cite{Snow2021}. 
However, their model neglects the role of radiative ionisation and recombination, and likely over-predicts the energy lost due to ionisation by assuming that all ionisation occurs from the ground state.


A more suitable model for the solar atmosphere involves solving the ionisation, recombination, excitation, and de-excitation rates for a multi-level model using both collisional and radiative rates. An underlying model is described by \cite{Sollum1999}, which is implemented in a single-fluid framework as part of the Bifrost code \citep{Leenaarts2007}. Their solar-like single-fluid simulations mimic many aspects of the solar chromosphere and have been used to study the heating consequences of shocks in the solar chromosphere \citep{Martinez2020}. However, a missing element is an understanding of the substructure of partially ionised shocks and how this contributes to the general shock behaviour, which can only be studied in a two-fluid framework.

Here we follow the Sollum method to implement a six-level (five neutral plus continuum/charged) hydrogen model into the two-fluid framework of the (P\underline{I}P) code, and use it to study switch-off slow-mode shocks. 
We self-consistently calculate the energy losses (due to ionisation and excitation) and gains (through the work done on the electron during recombination and de-excitation) which can act as sources/sinks of energy in our model. 1.5D simulations are performed using typical properties at different heights in the atmosphere. It is found that these partially ionised shocks have significantly different jump conditions to their MHD analogues. In particular, the partially-ionised shocks have enhanced density jumps and much less heating than their MHD analogues, which may have strong consequences for the role of shocks in heating the lower solar atmosphere.


\section{Shocks in partially ionised plasmas - the basics}

\subsection{Shock definitions}

An infinitesimally small disturbance in a fluid triggers waves that propagate at the characteristic wave speed of the system, allowing the upstream material to react such that the fluid response to the disturbance is smooth. For disturbances with a finite (i.e., non-infinitesimal) magnitude, different regions of the disturbance propagate with different speeds, some of which are faster than the characteristic wave speed of the background medium. This results in the upstream disturbances piling up on the propagating front, creating a sharp, steep jump. The stronger the disturbance the larger the jump becomes. This sharp fluid response is what is commonly known as a shock.

Explosive phenomena, such as magnetic reconnection, are known to drive shocks in the solar atmosphere, and models of fast magnetic reconnection often feature shocks \citep[e.g.,][]{Petschek1964,Liu2012,Innocenti2015,Shibayama2015}. 
In ideal MHD, there are three characteristic wave speeds (slow magnetoacoustic, Alfv\'en, fast magnetoacoustic) leading to a wealth of possible shock transitions, that can be broadly categorised into three types: slow, intermediate, and fast. To classify shocks, the fluid can be translated into the deHoffmann-Teller shock frame, where the shock is stationary and has zero electric field ($\textbf{v}\times \textbf{B}=\textbf{0}$). 
In the shock frame, the velocity normal to the shock front $v_\perp$, can be defined as existing in the following states either side of the shock:
\begin{itemize}
    \item (1) superfast - $V_{\rm fast} < v_\perp$
    \item (2) subfast - $V_{A} < v_\perp < V_{\rm fast}$
    \item (3) superslow - $V_{\rm slow} < v_\perp < V_{A}$
    \item (4) subslow - $0 < v_\perp < V_{\rm slow}$
    \item ($\infty$) static - $v_\perp = 0$
\end{itemize}
The permissible shock transitions are then defined as
\begin{itemize}
    \item Fast shock (from 1 to 2) 
    \item Slow shock (from 3 to 4)
    \item Intermediate shock (from 1 to 3, 1 to 4, 2 to 3, or 2 to 4)
\end{itemize}
A particular case is a switch-off slow-mode shock, which is a slow-mode shock with an inflow Alfv\'en Mach number ($v_\perp / v_A$) of unity, and the downstream magnetic field component parallel to the shock front is zero. The switch-off slow-mode shock is often produced by models of magnetic reconnection \citep[e.g., the Petschek-type shocks that form during the plasmoid instability,][]{Shibayama2015}. In this paper, the $2-4$ intermediate shock (i.e., a transition from above the Alfv\'en speed to below the slow speed) is also of interest since it is known to form within two-fluid switch-off shocks 
\citep{Snow2019}. 

\subsection{Shock jumps in partially ionsied plasmas}

For ideal MHD systems, a single equation can be derived that gives all possible shock transitions for a given upstream plasma-$\beta$ and angle of the magnetic field \citep{Hau1989}. In a conservative two-fluid, partially ionised system, one finds that sufficiently upstream and downstream of the shock, the MHD jumps must hold \citep{Snow2019}. This is understandable since sufficiently far from an isolated shock, both the ionsied and neutral species would be expected to have the same velocity (i.e., the drift velocity $\textbf{v}_{\rm p}-\textbf{v}_{\rm n} =0$) and the two-fluid equations should behave like a bulk MHD-like single-fluid. 

In a two-fluid model where collsional ionisation and recombination are considered, non-conservative terms appear in the plasma energy equation to account for the energy deposited by an electron to ionise a neutral atom, and the work done on a free electron during collisional recombination. For an idealised form of ionisation and recombination, \cite{Snow2021} showed both analytically and numerically that this non-conservative term significantly changes the stable jump conditions that exist across the shock, with shocks becoming much more compressible than the MHD limit. The ionisation losses can also become larger than the adiabatic heating resulting in net cooling across the shock interface. 

\subsection{Finite width of two-fluid shocks}

\begin{figure}
    \centering
    \includegraphics[width=0.95\linewidth]{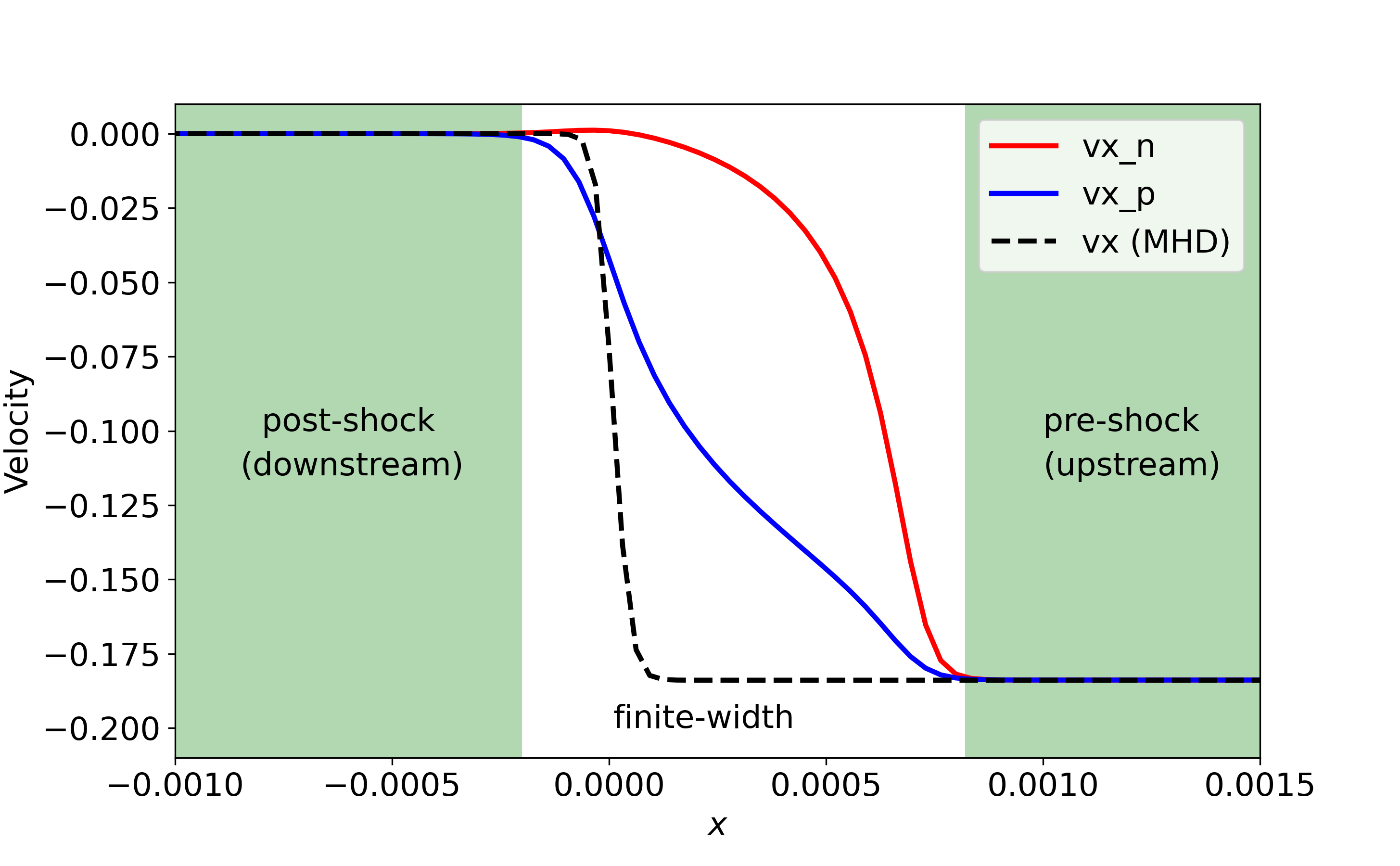}
    \caption{Switch-off slow-mode shock in a two-fluid model coupled through thermal collisions only (no ionisation/recombination). An MHD simulation is included for reference.}
    \label{fig:schemplot}
\end{figure}

The shock structure for a partially-ionised shock is significantly different to an idealised MHD shock. An idealised MHD shock is an effectively discontinuous jump in plasma properties across the shock front. In simulations, numerical/physical dissipation smooth the shock over a few grid cells preventing an infinite gradient existing. In the two-fluid model, the species separate and recouple around the shock front, as shown in Figure \ref{fig:schemplot} for a two-fluid simulation coupled using thermal collisions only (i.e., no ionisation and recombination). In both the upstream (pre-shock) and downstream (post-shock) regions, the two species are fully coupled (i.e., the drift velocity $v_{\rm p}-v_{\rm n} =0$) and reduce to the MHD state. However, a finite-width region exists where substructure forms due to the large drift velocities between the two fluids (i.e., the drift velocity $v_{\rm p}-v_{\rm n} \neq 0$). 

The substructure is known to host several interesting phenomena, including additional shock transitions \citep[e.g., sonic shock in the neutral species and an intermediate shock in the plasma,][]{Hillier2016,Snow2019} and cooling within the shock front itself \citep{Snow2021}. For two-fluid shocks that are coupled using thermal collisions only, the finite-width of the shock is known to scale with the ionisation fraction, i.e., the finite-width approaches zero as the ionisation fraction approaches unity \citep{Hillier2016}. In a stratified atmosphere, the width of the shock substructure can become comparable to the pressure scale height of the system \citep{Snow2020}.

\subsection{Limitations of previous work}

The ionisation and recombination model in \cite{Snow2021} has several strong assumptions that limit its applicability to the solar atmosphere. Firstly, the rates are empirically determined to fit temperatures above 1eV (11606 K), whereas the temperature of the solar chromosphere is typically $\approx 6000$ K. 
Secondly, the model only accounts for ground-state hydrogen whereas it is known that excited states of neutral hydrogen exist in the solar atmosphere and require less energy to ionise. Thirdly, the heating function was arbitrary and constant, whereas it should locally vary depending on the recombination rate. Finally, only collisional rates were included, whereas radiative rates are significant. 

The assumptions used in \cite{Snow2021} allowed for a simplified study to be performed, however limit the applicability of the results to the solar atmosphere. In this paper, all of these assumptions are relaxed, resulting in a far more appropriate underlying model for determining the behaviour of partially-ionised shocks in the lower solar atmosphere.


\section{Methods} \label{sec:methods}

We perform numerical simulations using the (P\underline{I}P) code \citep{Hillier2016} for a partially-ionised plasma. The two-fluid equations for a charge-neutral ion+electron fluid and a neutral fluid are given by:
\begin{gather}
\frac{\partial \rho _{\text{n}}}{\partial t} + \nabla \cdot (\rho _{\text{n}} \textbf{v}_{\text{n}})= \Gamma _{\rm rec} \rho _{\rm p} - \Gamma _{\rm ion} \rho _{\rm n}, \label{eqn:neutral1} 
\end{gather}
\begin{gather}
\frac{\partial}{\partial t}(\rho _{\text{n}} \textbf{v}_{\text{n}}) + \nabla \cdot (\rho _{\text{n}} \textbf{v}_{\text{n}} \textbf{v}_{\text{n}} + P_{\text{n}} \textbf{I}) \nonumber \\  \hspace{0.5cm} = -\alpha _c \rho_{\text{n}} \rho_{\text{p}} (\textbf{v}_{\text{n}}-\textbf{v}_{\text{p}}) + \Gamma _{\rm rec} \rho _{\rm p} \textbf{v}_{\rm p} - \Gamma _{\rm ion} \rho_{\rm n} \textbf{v}_{\rm n}, 
\end{gather}
\begin{gather}
\frac{\partial e_{\text{n}}}{\partial t} + \nabla \cdot \left[\textbf{v}_{\text{n}} (e_{\text{n}} +P_{\text{n}}) \right] \nonumber \\  \hspace{0.5cm}= -\alpha _c \rho _{\text{n}} \rho _{\text{p}} \left[ \frac{1}{2} (\textbf{v}_{\text{n}} ^2 - \textbf{v}_{\text{p}} ^2)+ \frac{1}{\gamma-1} \left(\frac{P_{\rm n}}{\rho_{\rm n}}-\frac{1}{2}\frac{P_{\rm p}}{\rho_{\rm p}}\right) \right] \nonumber \\ \hspace{0.5cm}+ \frac{1}{2} \left( \Gamma _{\rm rec} \rho _{\rm p} \textbf{v}_{\rm p} ^2 - \Gamma _{\rm ion} \rho _{\rm n} \textbf{v}_{\rm n} ^2 \right) 
+\frac{1}{ (\gamma-1)} \left( \frac{1}{2} \Gamma _{\rm rec} P_{\rm p} -\Gamma _{\rm ion} P_{\rm n} \right),
\end{gather}
\begin{gather}
e_{\text{n}} = \frac{P_{\text{n}}}{\gamma -1} + \frac{1}{2} \rho _{\text{n}} v_{\text{n}} ^2, \label{eqn:neutral2} \\
\frac{\partial \rho _{\text{p}}}{\partial t} + \nabla \cdot (\rho_{\text{p}} \textbf{v}_{\text{p}}) = - \Gamma _{\rm rec} \rho _{\rm p} + \Gamma _{\rm ion} \rho _{\rm n} \label{eqn:plasma1}\\
\frac{\partial}{\partial t} (\rho_{\text{p}} \textbf{v}_{\text{p}})+ \nabla \cdot \left( \rho_{\text{p}} \textbf{v}_{\text{p}} \textbf{v}_{\text{p}} + P_{\text{p}} \textbf{I} - \textbf{B B} + \frac{\textbf{B}^2}{2} \textbf{I} \right) \nonumber \\  \hspace{0.5cm}= \alpha _c \rho_{\text{n}} \rho_{\text{p}}(\textbf{v}_{\text{n}} - \textbf{v}_{\text{p}}) - \Gamma _{\rm rec} \rho _{\rm p} \textbf{v}_{\rm p} + \Gamma _{\rm ion} \rho_{\rm n} \textbf{v}_{\rm n},
\end{gather}
\begin{gather}
\frac{\partial}{\partial t} \left( e_{\text{p}} + \frac{\textbf{B}^2}{2} \right) + \nabla \cdot \left[ \textbf{v}_{\text{p}} ( e_{\text{p}} + P_{\text{p}}) -  (\textbf{v}_{\rm p} \times \textbf{B}) \times \textbf{B} \right] \nonumber \\  \hspace{0.5cm} =  \alpha _c \rho _{\text{n}} \rho _{\text{p}} \left[ \frac{1}{2} (\textbf{v}_{\text{n}} ^2 - \textbf{v}_{\text{p}} ^2)+ \frac{1}{\gamma -1} \left(\frac{P_{\rm n}}{\rho_{\rm n}}-\frac{1}{2}\frac{P_{\rm p}}{\rho_{\rm p}}\right) \right] \nonumber \\ \hspace{0.5cm}- \frac{1}{2} \left( \Gamma _{\rm rec} \rho _{\rm p} \textbf{v}_{\rm p} ^2 - \Gamma _{\rm ion} \rho _{\rm n} \textbf{v}_{\rm n} ^2 \right) -\frac{1}{ (\gamma-1)} \left( \frac{1}{2} \Gamma _{\rm rec} P_{\rm p} -\Gamma _{\rm ion} P_{\rm n} \right) \nonumber \\  \hspace{0.5cm}- \phi_I + \phi_{R}, \label{eqn:ep}
\end{gather}
\begin{gather}
\frac{\partial \textbf{B}}{\partial t} - \nabla \times (\textbf{v}_{\text{p}} \times \textbf{B}) = 0, \\
e_{\text{p}} = \frac{P_{\text{p}}}{\gamma -1} + \frac{1}{2} \rho _{\text{p}} v_{\text{p}} ^2, \\
\nabla \cdot \textbf{B} = 0,\label{eqn:plasma2}
\end{gather}
for a charge neutral plasma (subscript $\mbox{p}$) and neutral (subscript $\mbox{n}$) species. The fluid properties are given by density $\rho$, pressure $P$, velocity $\textbf{v}$, magnetic field $\textbf{B}$ and thermal+kinetic energy $e$. Both species follow ideal gas laws for the non-dimensional temperature $T$, namely $T_{\rm n} = \gamma P_{\rm n}/\rho_{\rm n}$ and $T_{\rm p} = \frac{1}{2} \gamma P_{\rm p}/\rho_{\rm p}$, where the specific gas ratio $\gamma=5/3$. $\textbf{I}$ is the three dimensional identity matrix.

The species are thermally coupled through the collisional coefficient $\alpha_c$ which is calculated as:
\begin{gather}
    \alpha _c = \alpha _0 \sqrt{\frac{T_{\rm p}+T_{\rm n}}{2}} \sqrt{\frac{1}{T_{\rm init}}}.
\end{gather}
The factor of $\sqrt{\frac{1}{T_{\rm init}}}$ is to normalise the collisional coefficient using the initial temperature $T_{\rm init}$ such that $\alpha_c (t=0) = \alpha _0$.

\subsection{Ionisation and recombination model} \label{sec:ionmodel}

The ionisation and recombination terms given in Equations \ref{eqn:neutral1}-\ref{eqn:plasma2} represent the ensemble change of species from neutral to ionised and vice versa. Given that neutral species in the solar atmosphere are regularly excited to higher states, and that this has consequences for the energy lost through ionisation, it is of benefit to use a multi-level hydrogen model to more accurately calculate these ensemble rates. We adapt the six-level (five neutral + continuum/ionised) model presented by \cite{Sollum1999} for use within our two-fluid model. It is assumed that all neutral species behave as a single neutral fluid such that the system can be modelled in a two-fluid approach. All dimensional variables are marked with a hat ~$\hat{}$.

In this framework, the total recombination is given by the amount of material recombining to any neutral level from the ionisation species, i.e.,
\begin{gather}
    \Gamma_{\rm rec} \rho_{\rm p} = \rho_{\rm p} (\hat{C}_{\rm{p},1}+\hat{C}_{\rm{p},2}+\hat{C}_{\rm{p},3}+\hat{C}_{\rm{p},4}+\hat{C}_{\rm{p},5})/\hat{\Gamma} \nonumber \\
        \hspace{1.2cm}+ \rho_{\rm{p}} (\hat{R}_{\rm{p},1}+\hat{R}_{\rm{p},2}+\hat{R}_{\rm{p},3}+\hat{R}_{\rm{p},4}+\hat{R}_{\rm{p},5})/\hat{\Gamma}, \\
        \hspace{1.2cm} = \rho_{\rm p} (\hat{\Gamma}_{\rm rec,col} + \hat{\Gamma}_{\rm rec,rad})/\hat{\Gamma},
\end{gather}
where $\hat{C}_{ij}$ is the dimensional collisional rate and $\hat{R}_{ij}$ is the dimensional radiative rate to level $i$ from level $j$. Note that the collisional and radiative rates are functions of the local electron number density and temperature, and their form is given in full across Sections \ref{sec:colratsec} and \ref{sec:radratsec}, and Appendix \ref{app:ratecoef}. The terms $\hat{\Gamma}_{\rm rec,col}, \hat{\Gamma}_{\rm rec,rad}$ represent the ensemble recombination rate (in units of s$^{-1}$) due to collisions and radiation respectively and are dimensional. To normalise these rates to a non-dimensional value consistent with the underlying equations, a normalisation factor $\hat{\Gamma}$ is included of the form
\begin{gather}
    \hat{\Gamma}=\hat{\Gamma}_{\rm rec,col}(t=0)/\tau_{\rm IR},
\end{gather}
such that at time $t=0$, the collisional recombination rate is $\tau_{\rm IR}$ which is a free parameter and can be used to scale the initial rates relative to the collisional timescale, as used in \citep{Snow2021}. In this paper, we use $\tau_{IR}=10^{-5}$ meaning that recombination in the background medium occurs on timescales $10^5$ longer than collisions. 

Similarly, the total ionisation rate is then defined as the amount of material entering the ionised species from any neutral level, i.e.,
\begin{gather}
    \Gamma_{\rm ion} \rho_{\rm{n}} = (\rho_{\rm{n}1} \hat{C}_{1,\rm{p}}+\rho_{n2}\hat{C}_{2,\rm{p}}+\rho_{\rm{n}3}\hat{C}_{3,\rm{p}}+\rho_{\rm{n}4}\hat{C}_{4,\rm{p}}+\rho_{\rm{n}5}\hat{C}_{5,\rm{p}})/\hat{\Gamma} \nonumber \\  
    \hspace{1cm}+(\rho_{\rm{n}1}\hat{R}_{1,\rm{p}}+\rho_{\rm{n}2}\hat{R}_{2,\rm{p}}+\rho_{\rm{n}3} \hat{R}_{3,\rm{p}}+\rho_{\rm{n}4}\hat{R}_{4,\rm{p}}+\rho_{\rm{n}5}\hat{R}_{5,\rm{p}})/\hat{\Gamma}, \\
    \hspace{1cm} = \rho_{\rm{n}} (\hat{\Gamma}_{\rm ion,col} + \hat{\Gamma}_{\rm ion,rad})/\hat{\Gamma},
\end{gather}
where $\rho_{\rm{n}1}$ is ground-state hydrogen, $\rho_{\rm{n}2}$ is the 1st excited state, etc. The total neutral density is defined as $\rho_{\rm{n}} = \Sigma \rho_{\rm{n}i}$ for $i=1,5$. The dimensional terms $\hat{\Gamma}_{\rm ion,col}, \hat{\Gamma}_{\rm ion,rad}$ represent the ensemble ionisation rates due to collisions and radiation respectively. The ionisation rates are also normalised by the factor $\hat{\Gamma}$. It should be emphasised here that the model we use is for hydrogen only.


\subsection{Collisional rates} \label{sec:colratsec}



Following \cite{Leenaarts2007}, the model of \cite{Johnson1972} is used to calculate the dimensional collisional ionisation, recombination, excitation and de-excitation coefficients $\hat{C}_{\rm exc},\hat{C}_{\rm ion}$ (in units of m$^3$ s$^{-1}$), which are then used to calculate the dimensional collisional rates $\hat{C}$. The excitation rates from level $i$ to level $j$ are calculated as:
\begin{gather}
    \hat{C}_{i,j} = \frac{g_i}{g_j} \hat{n}_e \hat{C}_{\rm{exc},ij},
\end{gather}
for $i<j$. The de-excitation rates are given by:
\begin{gather}
    \hat{C}_{j,i} = \hat{n}_e \hat{C}_{\rm{exc},ij} \exp \left( {\frac{E_i-E_j}{k_B \hat{T}_e}} \right),
\end{gather}
where $g_n=2n^2$ is the level degeneracy, $E_n$ is the ionisation energy from level $n$ (see Table \ref{tab:ioneng}), and the Boltzmann constant $k_B=1.38064852\times 10^{-23} \mbox{m}^2 \mbox{kg} \mbox{\,s}^{-2} \mbox{K}^{-1}$. 
Collisional ionisation and recombination rates are then given by:
\begin{gather}
    \hat{C}_{i,p} = \hat{n}_e \hat{C}_{{\rm ion},i,{\rm p}} \exp \left( {-\frac{E_{\rm p} - E_i}{k_B \hat{T}_e}} \right), \\
    \hat{C}_{p,i} = \hat{C}_{i,{\rm p}} \left[\frac{n_i}{n_{\rm p}}\right]_{{\rm Saha}} ,
\end{gather}
where subscript $p$ is the ionised/contiuum level, and the Local Thermal Equilibrium (LTE) Saha population ratio is defined as:
\begin{gather}
    \left[\frac{n_i}{n_{\rm p}}\right]_{{\rm Saha}}=\hat{n}_e \frac{g_i}{g_{\rm p}}\left( \frac{2 \pi k_B \hat{T}_e m_e}{h^2} \right)^{-3/2}\exp\left(-{\frac{E_{\rm p}-E_i}{k_B \hat{T}_e}} \right). \label{eqn:saha}
\end{gather}
The dimensional collisional coefficients $\hat{C}_{\rm exc},\hat{C}_{\rm ion}$ depend on the electron number density $\hat{n}_e$ and electron temperature $\hat{T}_e$. We assume that the electron number density is equal to the ion number density. The collisional excitation and ionisation coefficients $\hat{C}_{\rm exc},\hat{C}_{\rm ion}$ are defined in \cite{Johnson1972} and also included in Appendix \ref{app:ratecoef} for completeness.



The total collisional ionisation and recombination rates for the bulk neutral fluid and plasma are given by
\begin{gather}
    \hat{\Gamma}_{\rm ion,col} = \frac{1}{\hat{n}_n} \sum_{i=1}^{5} \hat{n}_i \hat{C}_{i{\rm p}} = \frac{\hat{n}_{\rm p}}{\hat{n}_{\rm n}} F(\hat{T}_e,\hat{n}_i), \\
    \hat{\Gamma}_{\rm rec,col} = \sum_{i=1}^{5} \hat{C}_{{\rm p}i}=\hat{n}_p^2 G(\hat{T}_e),
\end{gather}
where $\hat{\Gamma}_{\rm ion,col},\hat{\Gamma}_{\rm rec,col}$ are the bulk dimensional ionisation and recombination rates due to collisions only.

\subsection{Radiative model} \label{sec:radratsec}

The radiative model outlined in \cite{Sollum1999} is also implemented in our model for hydrogen. This consists of a background radiative field that is assumed to be blackbody. The dimensional radiative excitation and de-excitation coefficients take the form:
\begin{gather}
    \hat{R}_{lu}=\frac{4 \pi}{h \nu_0} \frac{\pi e^2}{m_e c}f_{lu}\frac{2 h \nu_0^3}{c^2} \frac{1}{e^{h \nu_0/k_B \hat{T}_{\rm rad}}-1},\\
    \hat{R}_{ul}=\frac{g_l}{g_u} e^{h \nu_0/k_B \hat{T}_{\rm rad}} R_{lu},
\end{gather}
with Planck's constant $h=6.62607004\times 10^{-34}$ m$^2$ kg s$^{-1}$, speed of light $c=299792458$ m s$^{-1}$, and charge of an electron $e=-1.6\times 10^{-19}$ Coulombs. The oscillator strengths $f_{lu}$ are given in Table \ref{tab:oscst} and are taken from \cite{Goldwire1968}.

The ionisation and recombination rates are given by:
\begin{gather}
    \hat{R}_{ip}=\frac{8 \pi}{c^2} \alpha_0 \nu_0^3 \sum^\infty _{j=1} E_1\left( \frac{j \nu_0}{k_B \hat{T}_{\rm rad}}\right), \label{eqn:radion}\\
    \hat{R}_{pi}=\frac{8 \pi}{c^2} \alpha_0 \nu_0^3 \left[ \frac{\hat{n}_i}{\hat{n}_p} \right]_{\rm LTE}  \sum^\infty _{j=1} E_1\left((j \hat{T}_e/\hat{T}_{\rm rad}+1) \frac{h \nu_0}{k_B \hat{T}_e}\right), \label{eqn:radrec}
\end{gather}
where $E_1(x)$ is the first exponential integral from \cite[p.78]{Rutten2003}, defined as
\begin{gather}
    E_1(x)=\int _1 ^\infty e^{-x \omega} \omega ^{-1} d\omega .
\end{gather}
The sum to infinity terms in Equations (\ref{eqn:radion})-(\ref{eqn:radrec}) are numerically taxing to solve during the simulation. Instead, look-up tables are created that cover the expected parameter range of the simulation. 

The radiative temperature is assumed to be equal to the local neutral temperature for the Lyman transitions and constant as $\hat{T}_{\rm rad}= 6000$ K for the other transitions. i.e. we assume the wavelengths of the Lyman transitions are sufficiently optically thick that mean free paths are small and so the the radiation temperature becomes approximately determined by the local plasma conditions. However, the wavelengths of light from other transitions have radiation temperatures
principally determined by the light emerging from below. It should be acknowledged that this is a rough approximation and the radiation temperature values for these transitions will deviate greatly through the varied, non-equilibrium conditions of the lower solar atmosphere.

\subsection{Ionisation/excitation energy loss, and recombination/de-excitation gain}

\begin{table}
    \centering
    \caption{Ionisation energy of hydrogen}
    \begin{tabular}{c | c c}
        Ionisation energy from state i & [eV] & [joules] \\
        \hline 
        $E_1$ & 13.6 & 2.1789921e-18  
        \\$E_2$ & 3.4 & 5.4474801e-19 
        \\$E_3$ & 1.51 & 2.4193221e-19  
        \\$E_4$ & 0.85 & 1.3618700e-19  
        \\$E_5$ & 0.54 & 8.6518801e-20 
    \end{tabular}
    \label{tab:ioneng}
\end{table}

During collisional ionisation, ambient thermal energy of the plasma is expended by a free electron to release a bound electron. Similarly, thermal energy is expended from the plasma to collisionally excite a neutral species to a higher energy state. 
As such, in our model, there is a macroscopic thermal energy loss from the plasma species due to collisional ionisation and excitation. This energy is calculated as
\begin{gather}
    \hat{I}_{\rm ion}= \Sigma \hat{n}_i \hat{C}_{i{\rm p}} E_i, \\
    \hat{I}_{\rm exc}= \Sigma \hat{n}_1 \hat{C}_{1u} (E_u-E_1) + \Sigma \hat{n}_2 \hat{C}_{2u} (E_u-E_2) \\
    \hspace{1cm} + \Sigma \hat{n}_3 \hat{C}_{3u} (E_u-E_3) + \hat{n}_4 \hat{C}_{4,5} (E_5-E_4),  \\
    \phi_{I}=(\hat{I}_{\rm{ion}}+\hat{I}_{\rm{exc}})/\hat{\phi} ,
\end{gather}
where $\hat{\phi}$ is a normalisation constant. The ionisation energy for each level is included in Table \ref{tab:ioneng}.

Similarly, an energy source exists that is the work done on the electron through collisional recombination and de-excitation processes that has the form
\begin{gather}
    \hat{R}_{\rm rec}= \Sigma \hat{n}_{\rm p} \hat{C}_{{\rm p}i} E_i ,\\
    \hat{R}_{\rm dex}= \Sigma \hat{n}_5 \hat{C}_{5l} (E_l-E_5) + \Sigma \hat{n}_4 \hat{C}_{l4} (E_l-E_4) \\
    \hspace{1cm} + \Sigma \hat{n}_3 \hat{C}_{l3} (E_3-E_l) + \hat{n}_2 \hat{C}_{2,1} (E_1-E_2) , \\
    \phi_{R}=(\hat{R}_{\rm rec}+\hat{R}_{\rm dex})/\hat{\phi} .
\end{gather}
This model is fully self-consistent, in that the heating mechanism is physically calculated from the rates, rather than being an arbitrary function. 
The definition of the rates is such that the equilibrium solution is an LTE Saha equilibrium and is in collisional equilibrium, that is $\phi_I=\phi_R$. As the simulation evolves, the losses and gains evolve leading to areas of net heating or cooling.

Note that these losses directly depend on the collisional rates only. However, the radiative rates also play a role in determining the ambient level populations on which these collisions act. Therefore, the radiation field indirectly impacts this source of heating or cooling. 
For example, collisional ionisation followed by radiative recombination causes a net energy decrease in the ambient plasma. 
This allows channels for the radiative field to contribute towards cooling, e.g., the radiative cooling of the solar corona. 
This model neglects the contribution of the high-energy tail during photo-ionisation/recombination, which can contribute towards heating/cooling and will be evaluated in future research.










\section{Implementation into (P\underline{I}P) code}

The atmosphere in our two-fluid model is assumed to be entirely hydrogen. 
It is also assumed that the electron temperature is equal to the plasma temperature, and that quasi-neutrality holds such that the local electron number density is equal to the plasma number density. All non-ionised levels of hydrogen are assumed to behave as a single neutral fluid (with the same velocity) and thus our two-fluid system models the ensemble neutral hydrogen (in any of the five excitation states), and a plasma species (consisting of ionised hydrogen and electrons). 


The neutral hydrogen level populations are evolved at each time step according to the rate equations:
\begin{gather}
    \frac{\partial n_i}{\partial t} = \Sigma _{i\neq j} (C_{ji}+R_{ji}) n_j - n_i \Sigma _{i\neq j} C_{ij}+R_{ij} - \textbf{v}_{n}\cdot \nabla (n_i) , \label{eqn:popevo}
\end{gather}
for hydrogen level $i$. For the convective term $\textbf{v}\cdot \nabla (n)$, all neutral levels ($i=1:5$) are assumed to use the neutral velocity $\textbf{v}_n$. A check is performed to ensure that the sum of the neutral excited levels $\Sigma _{i=1,5} n_i$ is equal to the density of the bulk neutral species $\rho_{\rm n}$. Any discrepancy is added/subtracted to the ground state neutral population which is the most populous neutral level in our simulations. The level populations are stored in double precision and a minimum simulation level population of $10^{-16}$ is enforced for numerical stability. For reference, the smallest level population produced in our simulations is $\approx10^{-13}$. 


\subsection{Energy in the system}

There are three broad types of energy in our model: macroscopic energy, ionisation/excitation energy, and radiation energy. The macroscopic energy (i.e., thermal+kinetic+magnetic) is the only energy directly modelled by the two-fluid equations used in this paper, and hence the term 'non-conservative' is used to describe a conversion of energy from this modelled macroscopic energy to the unmodelled ionisation/excitation energy, or a loss of energy through radiation. The ionisation/excitation energy is the energy stored in the proton-electron connection of a neutral particle. This is not directly modelled, however can be calculated from the neutral populations, i.e., by knowing the level populations of excited neutral states, the energy required for these states can be calculated. Since the macroscopic energy can be affected by collisional ionisation and recombination, ionisation/excitation energy can lead to macroscopic heating or cooling directly via work done on, or by, a free electron.

Radiation energy is supplied from an external field that adds ionisation/excitation energy. As such, radiation does not directly lead to heating or cooling in the macroscopic fluid, but can provide the ionisation/excitation energy, which can then indirectly lead to macroscopic heating through collisional processes involving the free electron. This is classified as indirect heating or cooling. Radiation energy is also lost from the ionisation/excitation energy in spontaneous emissions and stimulated de-excitations (also called "negative absorptions") and thus, similarly, can indirectly result in cooling of the macroscopic fluid through collisions. Note that the radiative model used in this paper is approximate and has several strong assumptions, as discussed in Section \ref{sec:radratsec}.




\section{Initial conditions} \label{sec:inic}

\begin{figure}
    \centering
    \includegraphics[width=0.95\linewidth]{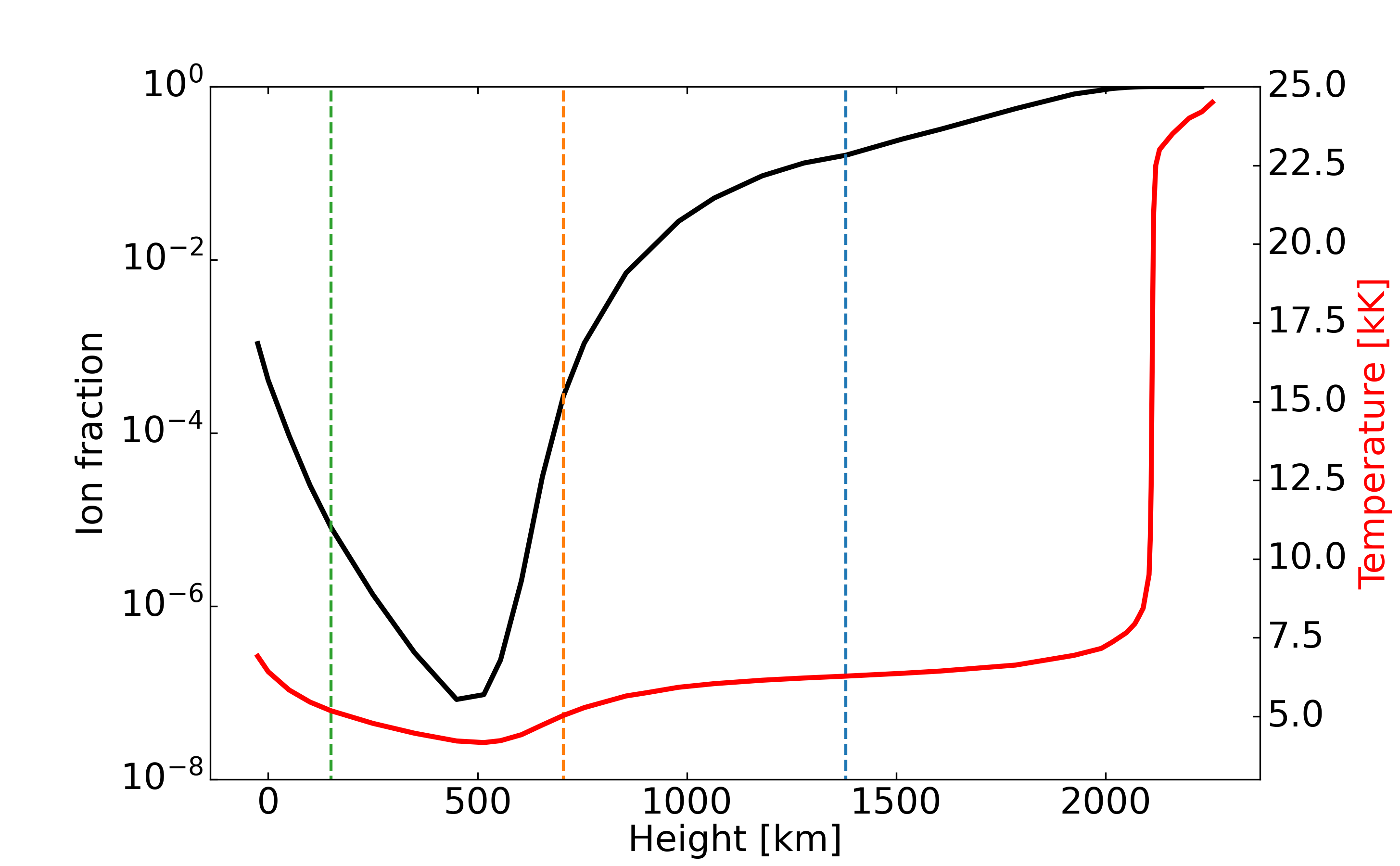}
    \caption{Ion fraction (black) and temperature in Kelvin (red) from the Saha equilibrium using the VALC temperature and density information. Vertical lines indicate the sampled heights for our numerical simulations.}
    \label{fig:saha}
\end{figure}

\begin{figure*}
    \centering
    \includegraphics[width=0.95\linewidth,clip=true,trim=0.9cm 0.8cm 0.9cm 0.8cm]{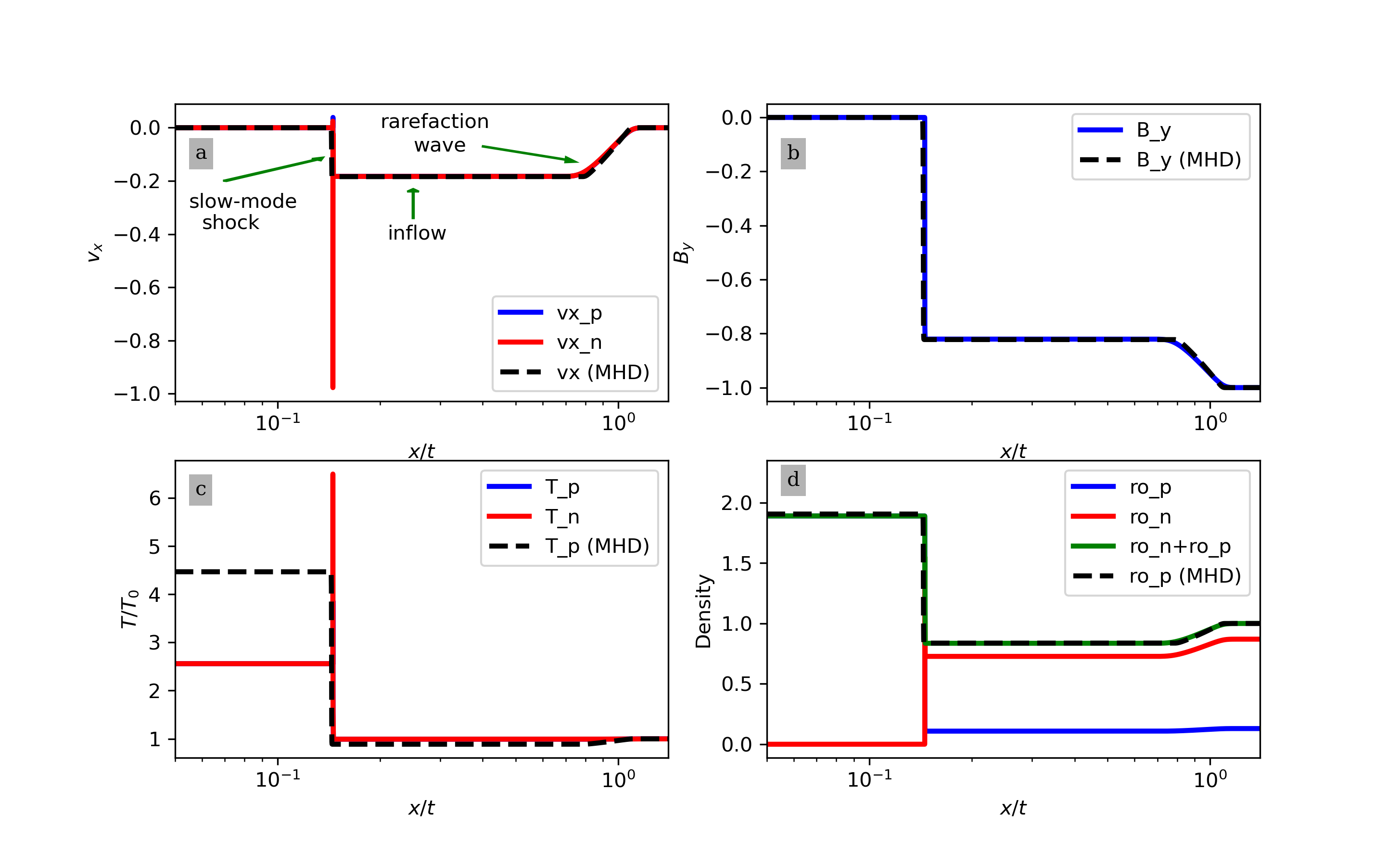}
    \caption{Upper-chromosphere case showing the $v_x$ velocity (top left), $B_y$ magnetic field (top right), temperature (lower left) and density (lower right). An MHD simulation is included for reference as the black line. For the PIP simulations, the plasma component is shown in blue and the neutral component in red. In the density plot, the green line shows the total ($\rho_n+\rho_p$) density in the PIP simulation. For panel c, the reference temperature $T_0=6220$.}
    \label{fig:upperchromocontext}
\end{figure*}

1.5D numerical simulations of two-fluid, switch-off slow-mode shocks are performed \citep{Hillier2016,Snow2019,Snow2021}. The initial conditions are characterised by a discontinuous magnetic field across the origin, with the plasma and neutral species in thermal equilibrium:
\begin{gather}
B_x = 0.1,
\end{gather}
\begin{gather}
B_y = -1.0 ~(x>0), 1.0 ~ (x<0), \\
\rho _{\text{n}} = \xi _{\text{n}} \rho _{\text{t}} ,\\
\rho _{\text{p}} = \xi _i \rho _{\text{t}} \textbf{=} (1- \xi _{\text{n}}) \rho _{\text{t}}, \\
P_{\text{n}} = \frac{\xi _{\text{n}}}{\xi_{\text{n}} + 2 \xi _i} P_{\text{t}} =  \frac{\xi _{\text{n}}}{\xi_{\text{n}} + 2 \xi _i} \beta \frac{B_0 ^2}{2} ,\\
P_{\text{p}} = \frac{2 \xi _i}{\xi_{\text{n}} + 2 \xi _i} P_{\text{t}} =  \frac{2 \xi _i}{\xi_{\text{n}} + 2 \xi _i} \beta \frac{B_0 ^2}{2},
\end{gather}
where $\xi _n$ and $\xi_i$ are the neutral and ion fractions respectively. The subscript $t$ denotes a total (plasma+neutral) quantity. Initially, plasma-$\beta=0.1$ and all velocity components are zero. The initial neutral fraction and population levels are calculated by iterating the population level equation (Equation \ref{eqn:popevo}) using the collisional and radiative rates calculated from the reference electron number density and electron temperature until convergence. A first-order HLLD solver is used to prevent spurious oscillations occurring around the shock interface. Note that for the 1.5D simulations studied here, the $\nabla \cdot \textbf{B}=0$ condition is trivially satisfied since $B_x=\mbox{const}$.


The system is normalised to a bulk density of unity ($\rho_{t}=\rho_n+\rho_p=1$), with the collisional frequency determined by the fluid density $\nu=\alpha_c (T_0) \rho_{t}=1$.  
The magnetic field is normalised as $B_0=B_{\rm norm}/\sqrt{4 \pi}$ leading to a bulk Alfv\'en speed $v_{A,t}=B_0/\sqrt{(\rho_n+\rho_p)}=1$. A length scale can be determined as $L_{\rm norm}=v_{A,t}/\nu$ which is the approximate length over which an Alfv\'en wave would become collisionally coupled i.e., the neutrals respond to an Alfv\'en disturbance. A reference electron temperature and electron number density are required since these determine the ionisation/recombination rates and the level populations. From the normalisation, a simulation plasma density $\rho_p=\xi_i$ corresponds to the reference electron number density $n_e$. 
The background recombination rate is normalised to $10^{-5}$, i.e., in the background medium, recombination occurs on timescales of $10^5$, whereas collisions occur on simulation timescales of $1$. 

As the simulation evolves, the magnetic field relaxes, triggering a fast-mode rarefaction wave that drives inflow towards a switch-off slow-mode shock. These initial conditions have been well studied in MHD and PIP using thermal collisions and empirical ionisation/recombination rates \citep{Hillier2016,Snow2019,Snow2021}. In the absence of loss/heating terms, the system behaves MHD-like sufficiently upstream and downstream of the shock, which can be proved analytically for any conservative effect (i.e., an effect that conserves total energy in the system) by studying the shock jumps \citep[see discussion in ][]{Snow2021}. The key differences arise within the shock front, which in the PIP models, has a finite-width determined by the coupling between ionised and neutral species. However, the results fundamentally change when losses/heating are included in the model since the plasma energy equation is no longer conservative (energy can now enter or leave the system due to ionisation losses or recombination gains) and hence the shock-jump equations are different. 

\subsection{Sampled heights}

Our initial electron number density and temperature (which also determine the initial neutral fraction) are taken from VALCIII model \citep{Vernazza1981}. Note that only the electron number density is taken from the VALCIII model, which is assumed equal to the plasma number density. The total density is calculated as $n_{t}=n_e/(1-\xi_n)$. The ionisation and recombination rates reach a SAHA-like LTE equilibrium as time tends towards infinity. As such, the initial ionisation fraction $\xi_i$ can be estimated using the VALC temperature and density as inputs to the Saha equation (Equation \ref{eqn:saha}). The resulting ionisation fraction as a function of height is shown in Figure \ref{fig:saha} alongside the temperature. Three atmospheric heights are chosen, which are indicated in Figure \ref{fig:saha} by the vertical dashed lines. These points sample very different regions of the chromosphere, with collisions dominating the lower chromospheric case, and radiation becomes more important in the upper chromosphere as the medium becomes more tenuous and collisions are less frequent. 

\begin{table}
    \centering
    \caption{Reference temperature, electron number density and neutral fraction for the sampled heights. The temperature and electron number density are from the VALCIII model \protect\citep{Vernazza1981}. Note that the electron number density is the same for the mid and upper chromospheric cases however the ion fraction and the total density are different. The plasma$-\beta$, collisional coefficient $\alpha _0$ and the reference recombination timescale $\tau_{IR}$ are constant across the simulations and are included here for completeness.}
    \begin{tabular}{c | c c c}
        Case & Upper & Mid & Lower \\
        \hline 
        $T_{\rm ref}$ [K] & 6220 & 5030 & 5180  
        \\$n_{\rm e,ref}$ [m$^{-3}$] & $7.5\times 10^{16}$ & $7.5\times 10^{16}$ & $6.5\times 10^{18}$ 
        \\$\xi_{\rm n}$ & 0.87016974 & 0.99973264 & 0.99999197
        \\plasma-$\beta$ & 0.1 & 0.1 & 0.1
        \\ $\alpha_0$ & 1 & 1 & 1
        \\ $\tau_{IR}$ & $10^{-5}$ & $10^{-5}$ & $10^{-5}$
    \end{tabular}
    \label{tab:sampledheights}
\end{table}

\begin{figure*}
    \centering
    \includegraphics[width=0.95\linewidth,clip=true,trim=1.2cm 0.8cm 0.9cm 0.8cm]{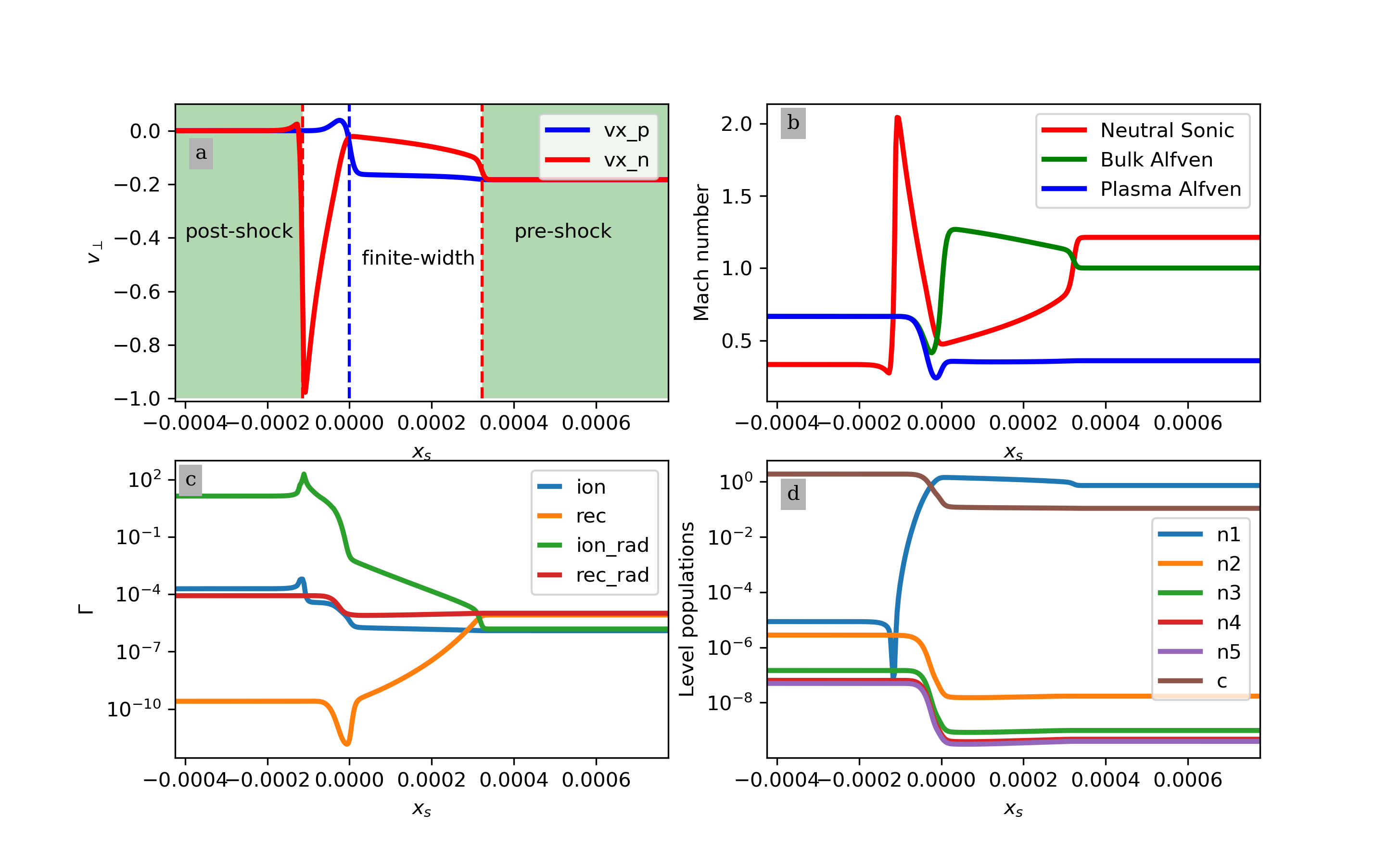}
    \caption{Shock substructure for the upper chromosphere case showing the shock velocity (top left), Mach numbers (top right), ionisation/recombination rates (lower left), and level populations (lower right). All plots are in the shock frame $x_s=x/t-v_s$. The vertical lines in the top left panel show the locations of the neutral sonic shock (red), and the plasma 2-4 intermediate shock (blue).}
    \label{fig:upchromosub}
\end{figure*}

\section{Upper-chromosphere case} \label{sec:upshock}

We first consider an upper-chromosphere case with reference temperature $T_0=6220$ K and reference electron number density $7.5\times 10^{16}$ m$^{-3}$, which results in an equilibrium neutral fraction of $\xi_{\rm n} \approx 0.870$.  
The neutral number density is $\approx 5.0\times 10^{17}$ m$^{-3}$, leading to a total number density of $\approx 5.8\times 10^{17}$ m$^{-3}$. The dimensional ion-neutral collisional frequency at time $t=0$ is calculated as 
\begin{gather}
    \nu_{\rm in}=n_{\rm n} \sqrt{\frac{8 k_B T}{\pi M}}\Sigma _{\rm in},
\end{gather}
with the mass of hydrogen $M=1.6735575\times 10^{-27}$ kg, and the ion-neutral collisional cross-section $\Sigma _{\rm in}=5 \times 10^{-19}$ m$^2$, which assumes equal temperatures of the ions and neutrals. Here this leads to an ion-neutral collisional frequency of $\nu_{\rm in}\approx 3\times 10^3$ Hz. The normalisation length scale $L_{norm}$ is calculated using the plasma-$\beta$ and the temperature as $L_{norm}=\frac{c_s}{\nu \sqrt{2 \gamma \beta}}$, where $c_s^2=\sqrt{\gamma k_B T/M}$. The description of the normalisation used for these simulations is given in Section \ref{sec:inic}.

The non-dimensional simulation is evolved for 6000 collisional times ($\approx 2$ seconds) using 512000 grid cells over a 1D domain with $x=[0,10000]$ ($\approx 48 \mbox{\,km}$) in order to resolve the shock substructure on the uniform grid. An MHD simulation is included for context which does not depend on the reference temperature and densities. The results are shown in Figure \ref{fig:upperchromocontext} in the quasi-self-similar $x/t$ frame such the shock is effectively stationary at late times.

The initial conditions are known to trigger a fast-mode rarefaction wave that drives inflow towards a switch-off slow-mode shock, as shown in Figure \ref{fig:upperchromocontext}a. This general structure is present in both the MHD and PIP simulations, however, there are several differences between the simulations. 
In terms of velocity, magnetic field strength and bulk density ($\rho_n+\rho_p$), the upper chromospheric case matches the MHD results well. However, while the shock inflow temperature is roughly the same in both simulations (a slight difference occurs due to the expansive rarefaction wave), the postshock temperature is significantly cooler in the PIP case compared to the MHD simulation, with the MHD simulation producing postshock temperatures around $4.47T_0$ ($\approx 27773$ K), and the PIP simulation around $2.6 T_0$ ($\approx 15919$ K). 

From Figure \ref{fig:upperchromocontext}d, one can see from the neutral and ionised densities that the ionisation fraction has changed across the shock in the PIP simulation. In the pre-shock medium, the neutral fraction is $\xi_n \approx 0.87$ (i.e., the medium is mostly neutral), however in the post-shock region the neutral fraction has reduced to $\xi_n\approx 6 \times 10^{-6}$ (i.e., the medium is almost entirely ionised). 
This explains the similarity to the MHD result; the neutral fraction is so low that the bulk dynamics are dominated by the plasma. The post-shock temperature is the main difference here which can be explained through the ionisation loss term (discussed in Section \ref{sec:uppsubshock}), and the treatment of the plasma species in a partially ionised medium compared to fully-ionised MHD. 
As a thought-experiment, consider an entirely neutral system that is spontaneously ionised entirely by the radiative field. In our model, we treat this as an energy neutral process where the photon provides the exact energy required to ionise the atom, resulting in a free electron with zero energy, i.e., the internal energy of the system is identical in both the neutral and ionised states. In such a system, the total mass density and pressure remain the same, however, the thermal energy is now shared between twice the number of species (protons and electrons). We assume that the thermal energy is split equally between the ions and electrons. The temperature of the ions (or electrons) in the fully ionised medium therefore depends on half the bulk plasma pressure divided by the total plasma mass, and hence is half the fully neutral temperature. In the upper chromospheric simulation, we have a mostly neutral plasma becoming almost entirely ionised from the radiative field and hence some reduction of temperature is expected since temperature is not a conserved quantity. 

Despite the general MHD-like behavior of this simulation, once can see from Figures \ref{fig:upperchromocontext}a and \ref{fig:upperchromocontext}c that something significantly different is occurring within the shock front. Namely, the neutral temperature and velocity are significantly greater than the pre- and post-shock values within the finite width of the shock. Analysing these requires a deeper look at the shock substructure, discussed further in Section \ref{sec:uppsubshock}.

\subsection{Shock substructure} \label{sec:uppsubshock}


\begin{figure}
    \centering
    \includegraphics[width=0.95\linewidth,clip=true,trim=0.2cm 2.7cm 1.0cm 3.5cm]{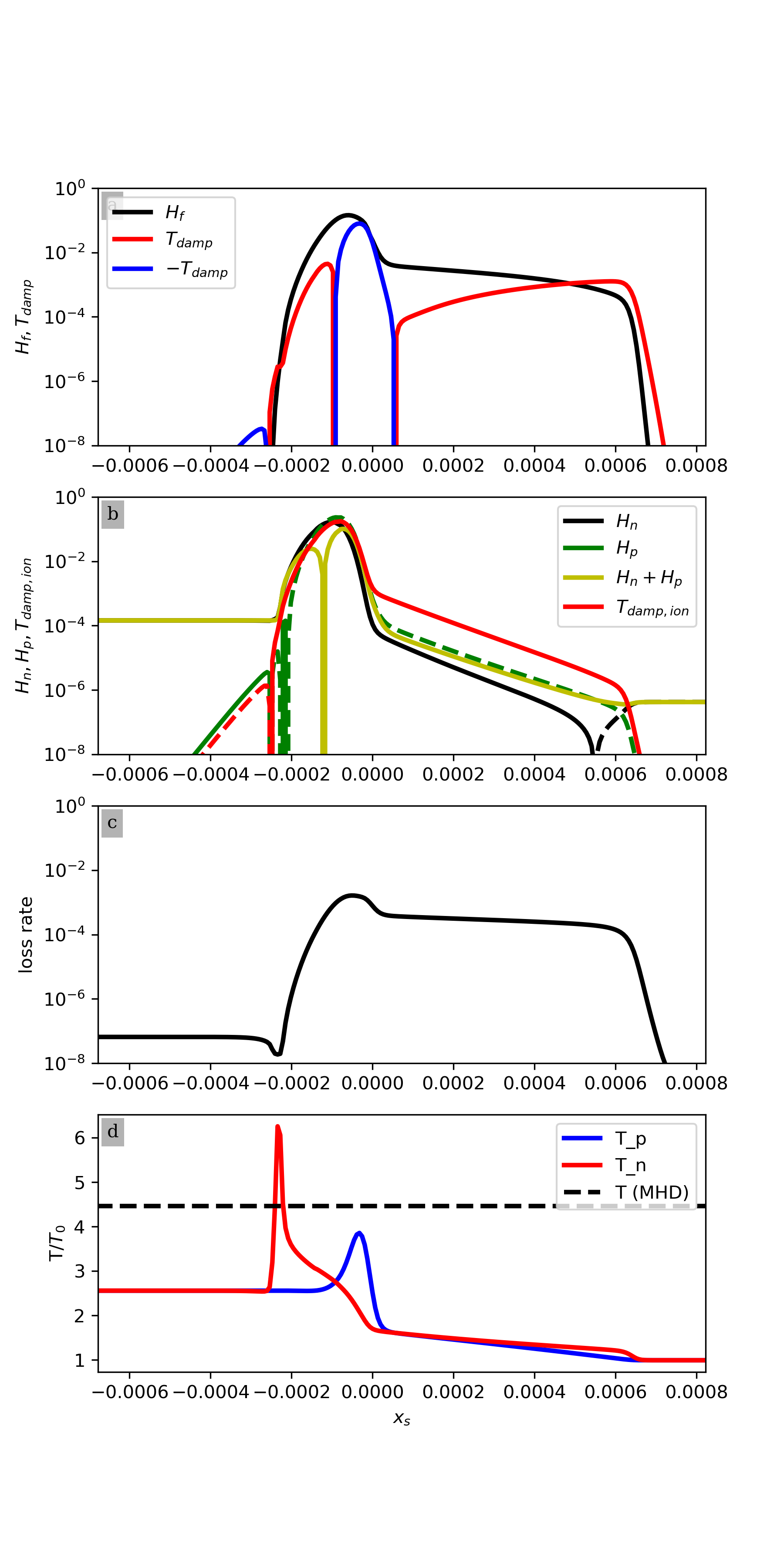}
    \caption{(a) frictional heating (black) and thermal damping across the shock for the upper chromosphere simulation. Red line indicates neutrals losing thermal energy to the plasma, and vice-versa for blue line. (b) frictional and thermal damping terms associated with ionisation and recombination. Negative values are denoted by a dashed line. (c) Net loss due to ionisation/recombination. (d) Temperature in the shock relative to the reference temperature $T_0=6220$ K. Black dashed line indicates the post-shock temperature in the MHD simulation.}
    \label{fig:upchromoheat}
\end{figure}

\begin{figure*}
    \centering
    \includegraphics[width=0.99\linewidth,clip=true,trim=0.0cm 0.0cm 0.0cm 0.0cm]{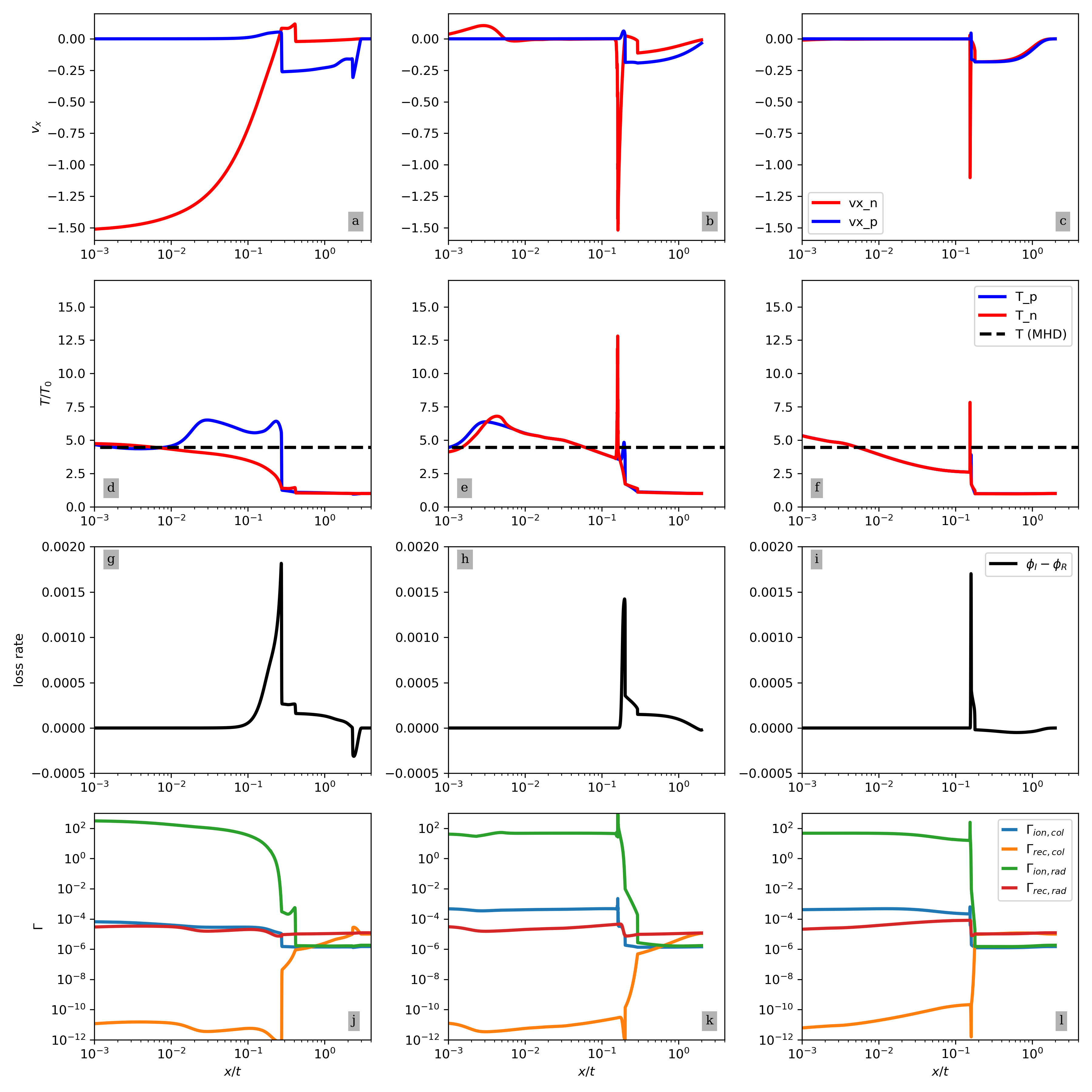}
    \caption{Time series for the upper chromosphere case at time $t=1$ (left column), $t=10$ (middle column), and $t=100$ (right column) collisional times, showing the velocity (panels a,b,c), temperature (panels d,e,f), net loss/heating (panels g,h,i), and ionisation/recombination rates (panels j,k,l). Plasma velocity and temperature are in blue, and neutrals in red. The post-shock temperature in the MHD simulation is plotted in panels d,e,f as the black dashed line. The rates are coded as $\Gamma_{\rm ion,col}$ (blue), $\Gamma_{\rm ion,rec}$ (green), $\Gamma_{\rm rec,col}$ (orange), $\Gamma_{\rm rec,rad}$ (red).}
    \label{fig:upchromots}
\end{figure*}

A key element of partially-ionised shocks is that the shock has a finite-width associated with the level of coupling between the two fluids. Figure \ref{fig:upchromosub} shows the substructure in the PIP simulation of the upper chromospheric case. Note that the axis here has been converted into a self-similar shock frame $x_s=x/t-v_s$ such that the shock is located at $x_s=0$, and the velocity $v_\perp = v_x-v_s$ where $v_s$ is the shock speed. 
Sufficiently upstream or downstream of the shock, the ions and neutrals are collisionally coupled (Figure \ref{fig:upchromosub}a) and in thermal balance (Figure \ref{fig:upchromoheat}c), defining the boundaries of our large-scale shock. 
Across the large-scale shock structure, the jump conditions indicate a switch-off slow-mode shock that has a transition from an inflow Alfv\'en Mach number of unity, to a sub-slow post-shock speed, and additional shock transitions occur within this larger-scale switch-off shock (Figure \ref{fig:upchromosub}b). 
The shock substructure consists of two main components: 1) a collisionally-decoupled shock; and 2) a post-shock neutral depletion.

On the leading edge of the shock, the neutrals separate from the plasma and produce a sonic shock (shown in Figure \ref{fig:upchromosub}a by the red dashed line, and Figure \ref{fig:upchromosub}b by the transition from super-sonic to sub-sonic) with an increase in ground state density (Figure \ref{fig:upchromosub}d) and temperature (Figures \ref{fig:upperchromocontext}c and \ref{fig:upchromoheat}d). The only bound state of hydrogen that increases in density here is the ground state (see Figure \ref{fig:upchromosub}d) since the temperature increase leads to efficient ionisation of the excited states and thus their level populations decrease slightly at the neutral shock at the leading edge of the shock. The temperature increase gradually couples to the plasma due to both thermal damping and ionisation. 

Following the neutral shock, the plasma accelerates relative to the local Alfv\'en speed. The plasma then shocks (indicated by the blue dashed line in Figure \ref{fig:upchromosub}a) however here the transition is super-Alfv\'enic to sub-slow, indicating a 2-4 intermediate shock. 2-4 intermediate shocks have been found to be a common substructure in partially-ionised switch-off slow-mode shocks, with the magnitude of the field reversal decreasing with time \citep[as described in][]{Snow2019}. Here the field reversal is $B_y\approx 10^{-10}$ compared to the background value $B_0=1$ and hence is not visible in Figure \ref{fig:upperchromocontext}b which shows the magnetic field through the domain. The small field reversal is likely due to the efficient coupling between species and the long simulation time. 

Following the plasma shock, there is a post-shock neutral depletion region where the increase in ionisation rates (Figure \ref{fig:upchromosub}c) leads to rapid ionisation and excitation of the neutral hydrogen. 
The strong depletion of ground state hydrogen leads to a shock occurring in the neutrals, after the plasma shock. This shock however is very low density $\approx 10^{-5}$ and hence the coupling between the ions and neutrals is weak since the collisional exchange of momentum and energy depends on the product of the densities and the drift velocity is on the order of 1. As such, this region can only efficiently couple through ionisation and recombination.  

The ionisation/recombination rates change drastically across the shock, and within its substructure. Outside of the finite width, the medium is in an effective ionisation/recombination equilibrium, i.e., $\Gamma _{\rm ion} \rho_{\rm n}-\Gamma _{\rm rec} \rho_{\rm p}\approx 0$, where $\Gamma_{\rm ion},\Gamma_{\rm rec}$ include both the collisional and radiative rates. As shown by Figure \ref{fig:upchromosub}c, in the inflow (pre-shock) region, recombinative rates are greater than the ionisation rates, and hence the medium is mostly neutral, with both collisional and radiative rates approximately equal. Post-shock, the medium is dominated by ionisation, with the radiative rates being significantly larger than the collisional rates. 
The increase in collisional ionisation (blue line \ Figure \ref{fig:upchromosub}c) leads to a large increase in the energy loss from the plasma (since energy is expended by the free electron to release a bound electron). The plasma thermal energy expenditure to ionise the neutrals effectively cools the post-shock medium to far below the MHD value in both fluid (due to the thermal coupling). Within the shock, there is a large increase in radiative ionisation due to the high neutral temperature leading to an increase in the radiative Lyman transitions (due to the optically thick approximation model used). The strong ionisation at the plasma shock leads to a reduction in thermal pressure of the neutral species, creating a neutral pressure gradient force that acts in the negative $x$-direction.

In this simulation, the pre-shock region is dominated by ground-state hydrogen or plasma, as shown in Figure \ref{fig:upchromosub}d, with relatively low populations of excited states. Post-shock, the medium is dominated by plasma, with low populations of neutral and excited states of hydrogen. 
In the lower chromosphere, larger populations of excited states are expected and thus they should be more important in determining the energy loss and compression at lower atmospheric heights.

The interactions between ionised and neutral species through both thermal collisions and ionisation/recombination leads to several heating and thermal damping terms, that are derived in Appendix \ref{app:heatdamp}. The thermal collisions lead to frictional heating ($H_{f}$) and thermal damping ($T_{\rm damp}$) between the species that are defined as
\begin{gather}
    H_{f}=\frac{1}{2}\alpha_c \rho_{\rm n}\rho_{\rm p}(v_{\rm n}-v_{\rm p})^2 ,\\
    T_{\rm damp}=\frac{1}{\gamma (\gamma -1)}\alpha_c \rho_{\rm n} \rho_{\rm p} (T_{\rm n}-T_{\rm p}) .
\end{gather}
Similarly, heating and damping terms exist due to ionisation and recombination process that are defined as
\begin{gather}
    T_{\rm damp,ion}=\frac{1}{\gamma(\gamma -1)} (\Gamma_{\rm ion} \rho_{\rm n} T_{\rm n}- \Gamma_{\rm rec} \rho_{\rm p} T_{\rm p}) ,\\
    H_{\rm n}=\frac{1}{2}\left( \Gamma_{\rm rec}\rho_{\rm p}\textbf{v}_{\rm p}^2 -2\Gamma_{\rm rec} \rho_{\rm p} \textbf{v}_{\rm p} \cdot \textbf{v}_{\rm n} + \Gamma_{\rm ion} \rho_{\rm n} \textbf{v}_{\rm n}^2 \right) ,\\
    H_{\rm p}=\frac{1}{2}\left( \Gamma_{\rm rec}\rho_{\rm p}\textbf{v}_{\rm p}^2 -2\Gamma_{\rm ion} \rho_{\rm n} \textbf{v}_{\rm n} \cdot \textbf{v}_{\rm p} + \Gamma_{\rm ion} \rho_{\rm n} \textbf{v}_{\rm n}^2 \right),
\end{gather}
which are plotted in Figure \ref{fig:upchromoheat} alongside the net loss rate ($\phi_I-\phi_R$) and temperature. 
At the teading edge of the shock, the sonic shock in the neutrals increases the neutral temperature, leading to thermal coupling from the neutrals to the plasma, increasing the plasma temperature. The increase in plasma temperature results in an increase in ionisation, and hence an increase in the loss rate (Figure \ref{fig:upchromoheat}c), however this term is smaller than the frictional heating term (Figure \ref{fig:upchromoheat}a) and hence there is a gradual increase in temperature of both the ions and neutrals (Figure \ref{fig:upchromoheat}d). 

The thermal damping due to ionisation and recombination is always positive in the shock as there is net ionisation and thus the neutrals are transferring their thermal energy to the plasma through ionisation. 
In terms of the heating terms, $H_{\rm n}$ is always positive and $H_{\rm p}$ is mostly negative in the shock due to the net ionisation within the shock. 

Figure \ref{fig:upchromoheat}c shows the net loss rate $\phi_N=\phi _I-\phi _R$, i.e., the difference between the energy lost through ionisation/excitation and that gained by recombination/de-excitation. 
Within the shock, the net loss rate is always positive as a result of the net ionisation of the medium, with the collisional ionisation leading to ionisation losses. The radiative ionisation (which has no directly associated losses) is far larger than the collisiona ionisation, and hence the energy lost through the shock is not as large as it would be in a purely collisional plasma, note how the maximum plasma temperature within the shock is $\approx 80\%$ of the MHD post-shock temperature. Lower atmospheric heights are more collisionally dominated and hence may experience greater cooling within the shock. The losses are relatively low post-shock and hence the temperature is maintained in the post-shock medium for relatively long time scales. In the rarefaction wave, which is expansive and slightly reduces the ion fraction, the net loss rate is slightly negative leading to a slight heating of the system. In this simulation, the heating across the rarefaction wave is negligibly small and the losses across the shock are of far more interest. 






\begin{figure*}
    \centering
    \includegraphics[width=0.95\linewidth,clip=true,trim=0.9cm 0.8cm 0.9cm 0.8cm]{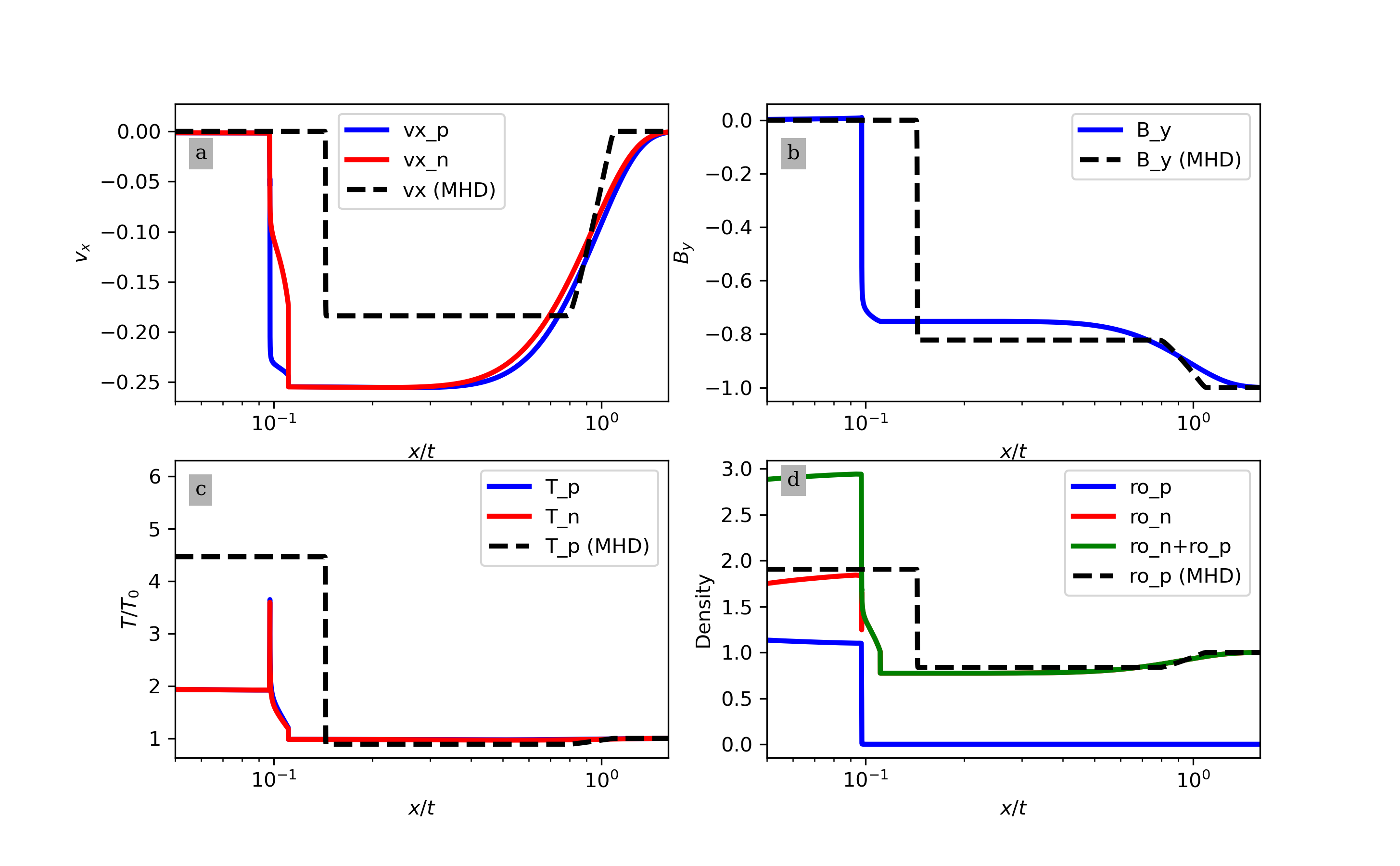}
    \caption{Context plot of the mid-chromosphere case ($T_0=5030$ K, $n_e=7.5\times 10^{16}$ m$^{-3}$, $\xi_n\approx 0.9997$) after 100,000 collisional times, showing the velocity (top left), perpendicular magnetic field (top right), temperature (lower left) and density (lower right). An MHD simulation is included for reference. 
    }
    \label{fig:midchromocontext}
\end{figure*}
\subsection{Early time evolution}

In the previous section, the results were analysed at a late time where the shock structure has reached a self-similar state. The early time evolution is also of interest since it tells us about the initiation phases of these reconnection-like shocks. Figure \ref{fig:upchromots} shows the simulation after 1, 10 and 100 collisional times in the quasi self-similar $x/t$ axis. Note that this axis is a function of time and hence structures that appear the same length are much larger at late times. 
Since the initial discontinuity is in the magnetic field only, it only directly drives the plasma component, which then couples to the neutrals through thermal collisions or recombination. At time $t=1$ collisional times (Figure \ref{fig:upchromots}, first column), the neutrals and plasma have very different velocity structures given the limited coupling possible in this time. The neutrals have reacted to the plasma but are not in a coupled state. The plasma temperature spikes at the shock front to greater than the MHD post-shock temperature. This is due to the plasma having an effectively lower plasma-$\beta$ (in our setup, the plasma$-\beta$ is based on the total pressures such that a fully coupled system with no losses would pair with the MHD simulation outside of the shock).

After 10 collisional times (Figure \ref{fig:upchromots}, middle column) one can see that there are some similar structures emerging in both the neutral and plasma species but they are still far from a steady state. Very high temperature of approximately $13T_0$ ($\approx 8 \times 10^4$ K) form in the neutral species at early times. This is mostly due to the rapid ionisation lowering the density of the neutral species increasing the efficiency of frictional heating and hence raising the neutral temperature. 

After 100 collisional times (Figure \ref{fig:upchromots}, right column), the velocity and temperature in the rarefaction wave and shock are becoming coupled, as evidenced by the smaller difference between the two fluids. The maximum plasma temperature inside the shock is now comparable to the MHD temperature, however the neutral temperature is much larger due to the rapid depletion of ground state hydrogen. The rates are dominated by radiative ionisation post-shock which is orders of magnitude larger than the pre-shock value. Radiative ionisation however has no direct associated loss in our model hence the peak losses do not change much throughout the simulation as these are determined by collisional rates only. Within the shock, the losses peak and result in efficient cooling. The losses only directly depend on collisional ionisation, and even though this is smaller than the radiative ionisation, the resultant losses consistently cool both the shock interior and downstream region. 

The normalisation of the system is such that initial background collisional recombination rate is $10^{-5}$ meaning that recombination occurs on timescales of $10^5$ collisional times. Within the shock, the ionisation (both collisional and radiative) increase greatly with the radiative ionisation increasing $~10$ orders of magnitude after 10 collisional times. As such, even though the background rates occur on relatively long timescales, within the shock the ionisation rates become significant on short time scales. This implies that at early times in the evolution of shocks, the dominant coupling mechanism is ionisation, rather than thermal collisions.


\section{Mid chromosphere case}

\begin{figure*}
    \centering
    \includegraphics[width=0.9\linewidth,clip=true,trim=1.2cm 0.8cm 0.9cm 0.8cm]{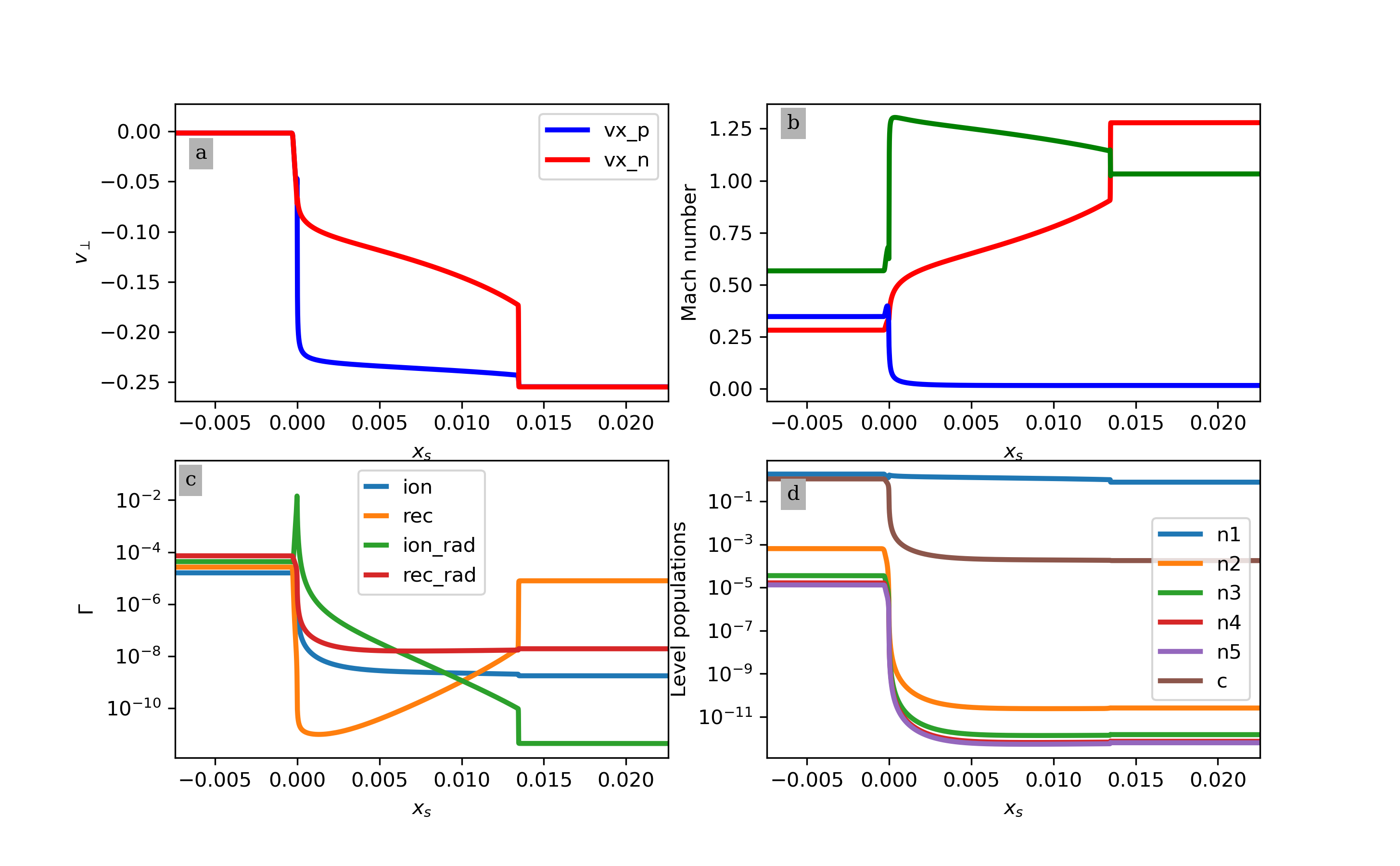}
    \caption{Shock substructure in the mid-chromospheric case showing the velocity (top left), Mach numbers (top right), collisional and radiative ionisation and recombination rates (lower left), and level populations (lower right). Note that this plot is in the shock frame. }
    \label{fig:midchromosub}
\end{figure*}

The next case considered is a mid-chromospheric case with a reference temperature of $T_0=5030$ K and electron number density $n_e=7.5\times 10^{16}$ m$^{-3}$, leading to an initial neutral fraction of $\xi_n\approx 0.9997$. In dimensional units, the ion number density is $7.5\times 10^{16}$ m$^{-3}$ (as with the upper chromospheric case), however since the neutral fraction has changed, the neutral number density is now $2.78\times 10^{20}$ m$^{-3}$. This leads to a dimensional ion-neutral collisional frequency as $\nu_{\rm in}=1.4\times 10^6$ Hz. 

Here the system is much more neutral than the upper chromospheric case and as such, a much longer simulation time is required to reach a quasi-self-similar solution. The non-dimensional simulation is evolved for 100,000 collisional times ($\approx 0.08$ seconds) on a domain $x=[0,2\times 10^5]$ ($\approx 2$\,km) using 512,000 grid cells.

Figure \ref{fig:midchromocontext} shows the large-scale behaviour. As before, both the PIP and MHD simulations have the same types of features (a fast-mode rarefaction wave that drives inflow towards a slow-mode shock) however the two-fluid simulation shows remarkably different quantitative behaviour. The two-fluid simulation has a much larger compression across the shock ($r\approx 3.8$) than the MHD case (Figure \ref{fig:midchromocontext}d), and a far cooler post-shock temperature ($T^d\approx 1.9 T_0 \approx 9768$ K), as shown in Figure \ref{fig:midchromocontext}c. The finite-width of the shock is also much larger here than in the upper chromosphere, and can clearly be seen in the velocity plot (Figure \ref{fig:midchromocontext}a). In the absence of ionisation and recombination, the finite-width of the shock is known to scale with the neutral fraction, with a larger neutral fraction resulting in a larger finite-width \citep{Hillier2016}. 

The shock substructure (Figure \ref{fig:midchromosub}) is far less structured than the upper-chromosphere case. Here, the finite-width of the shock consists solely of a collisionally decoupled region and there is additional shock forming after the plasma shock. This is because, as one moves lower in the atmosphere, the medium become more collisional and hence thermal collisions and collisional ionisation/recombination are the dominant effects. The radiative rates are smaller than their collisional counterparts in the pre-shock region, but become dominant post-shock (Figure \ref{fig:midchromosub}c). 

Within the finite shock width, the radiative ionisation increases and becomes dominant  but is of a comparable magnitude to the post-shock recombination rate (Figure \ref{fig:midchromosub}c). This is in contrast to the upper chromospheric case where the ionising radiation was many orders of magnitude larger than the other rates within the shock. Again, this behaviour can be attributed to the increased collisionality of the system that reduces the influence of the radiative field.

Despite the different behaviour, the shock transitions in the mid-chromospheric case are the same as in the upper chromospheric case, namely a large-scale switch-off slow-mode shock that hosts a sonic neutral shock and a 2-4 intermediate shock within its finite-width (Figure \ref{fig:midchromosub}b). The magnetic field reversal here is $B_y\approx0.01$, which is significantly larger than in the upper chromospheric case, and is visible in Figure \ref{fig:midchromocontext}b. The post-shock transverse magnetic field reversal rapidly reduces to zero, as is required by the large-scale switch-off slow-mode shock.

The level populations show that in the post-shock region, the excited states of neutral hydrogen have become significant and the post-shock $n_2$ population is comparable to the pre-shock plasma population (Figure \ref{fig:midchromosub}d). As such, a not-insignificant amount of energy has be lost due to excitation. Note that this is still much smaller than the energy lost due to ionisation from ground state, where most of the pre-shock material resides. 


The maximum temperature within the shock is lower than both the MHD and the upper chromosphere cases. The mid-chromospheric case is more collisionally dominated and hence a significant proportion of the energy required for ionisation comes from the free electrons, and hence the plasma. As such, there is a significant energy loss resulting in a lower post-shock temperature.  

The temperature jump across the shock (Figure \ref{fig:midchromocontext}c) is lower in the mid chromospheric case ($T_{\rm jump}=9768/5030\approx 2.0$) compared to the upper chromospheric case ($15919/6220\approx 2.6$). The mid-chromospheric case is significantly more collisional than the upper chromosphere. 
Even though the radiative ionisation rates are greater than the collisional ionisation rates through most of the shock, the ratio between these values is lower than the upper chromosperic case; in the mid-chromosphere, $\Gamma_{\rm ion}/\Gamma_{\rm ion\_rad} \approx 3.6 \times 10^{-3}$ at the at the location of maximum ionisation, compared to $\approx 3.2 \times 10^{-6}$ in the upper chromospheric case. As such, significantly more collisional ionisation occurs and hence more energy is lost through ionisation. 
Furthermore, in the mid-chromopsheric case, the medium remains mostly neutral post-shock (Figure \ref{fig:midchromosub}d), whereas in the upper-chromospheric case the post-shock region was mostly ionised.  


\section{Lower chromosphere case}


\begin{figure*}
    \centering
    \includegraphics[width=0.95\linewidth,clip=true,trim=0.9cm 0.8cm 0.9cm 0.8cm]{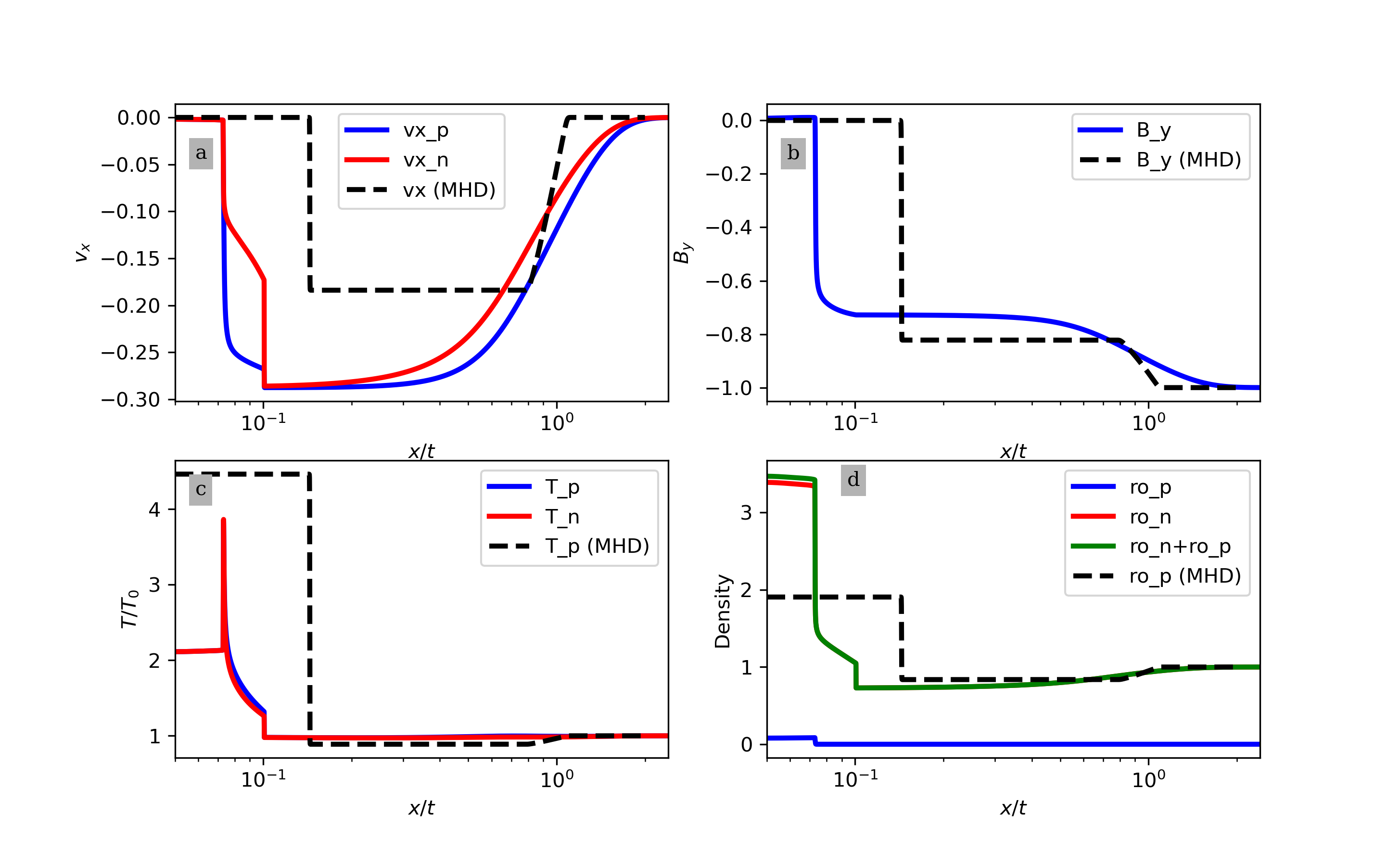}
    \caption{Context plot of the lower chromosphere case ($T_0=5180$ K, $n_e=6.5\times 10^{18}$ m$^{-3}$, $\xi_n\approx 0.999992$) after 1,000,000 collisional times. Layout is the same as in Figure \protect\ref{fig:upperchromocontext}}
    \label{fig:lowchromocontext}
\end{figure*}

\begin{figure*}
    \centering
    \includegraphics[width=0.9\linewidth,clip=true,trim=1.2cm 0.8cm 0.9cm 0.8cm]{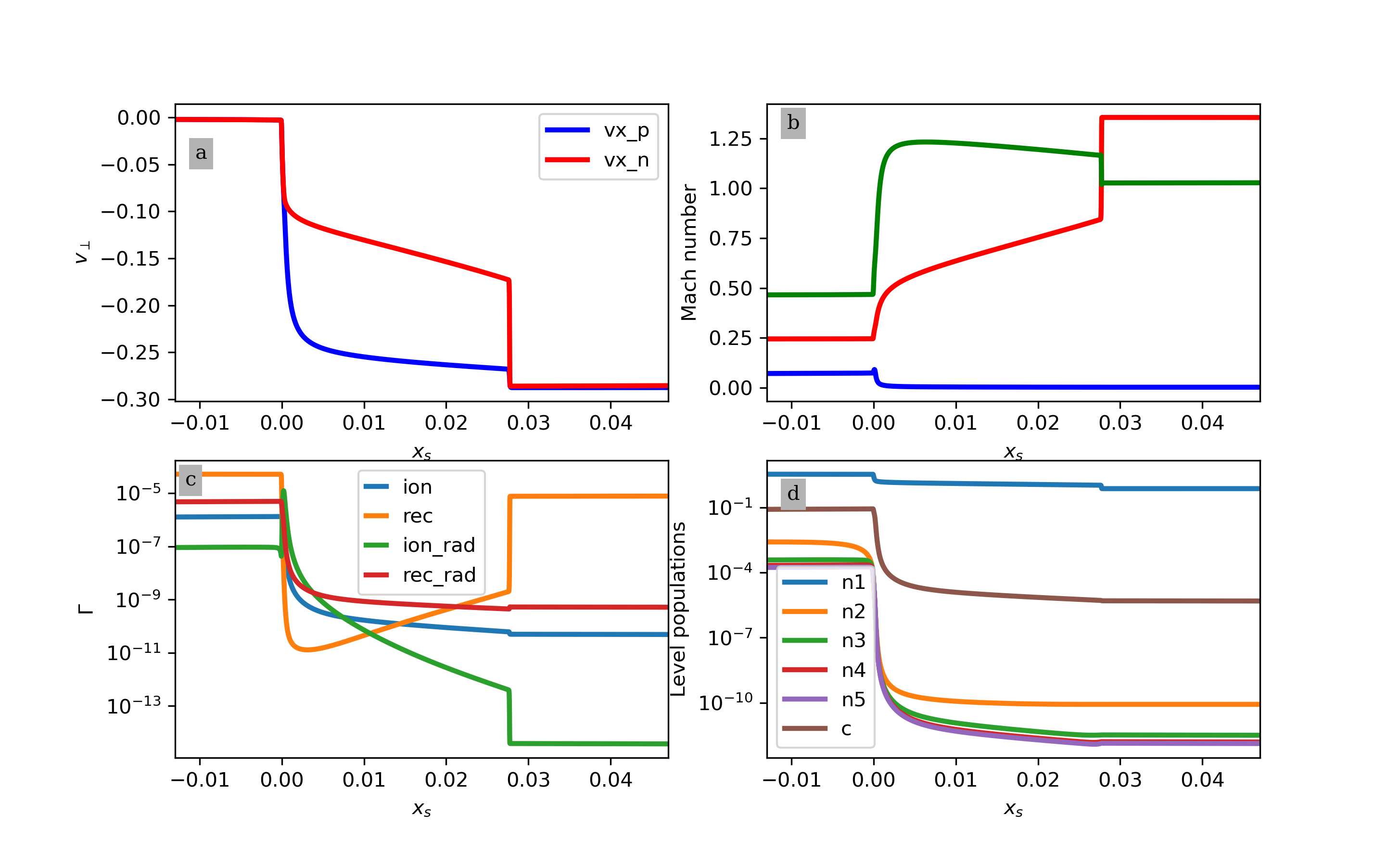}
    \caption{Shock substructure in the lower chromospheric case. Layout is the same as in Figure \protect\ref{fig:upchromosub}}
    \label{fig:lowchromosub}
\end{figure*}

The final atmospheric height considered is the lower chromosphere, which has a reference temperature of $T_0=5180$ K and a reference electron number density of $n_{\rm e,ref}=6.5 \times 10^{18} $ m$^{-3}$, leading to an initial neutral fraction of $\xi_{\rm n}\approx 0.999992$. This is significantly more neutral and much more dense than the previous heights considered. In dimensional units, the ion number density is $6.5\times 10^{18}$ m$^{-3}$ and the neutral number density is $8.12\times 10^{23}$ m$^{-3}$. This leads to a dimensional ion-neutral collisional frequency as 
$\nu_{\rm in}=4.2\times 10^9$ Hz. 
As such, the lower-chromospheric simulation is highly collisionally dominated. The non-dimensional simulation was evolved for 1,000,000 collisional times ($\approx 2.4 \times 10^{-4}$ seconds) on a grid spanning $x=[0, 10^6]$ ($\approx 70$\,m) using 1,024,000 grid cells.

Figure \ref{fig:lowchromocontext} shows the large-scale behaviour across the system. The key qualitative features remain consistent across all cases considered, namely that there is a rarefaction wave that drives inflow towards a slow-mode shock. Compared to the MHD simulation, there is significantly more compression across the shock ($r\approx 4.74$) and a significantly lower temperature jump ($T^d/T^u\approx 2.12/0.98\approx 2.17$) in the lower chromospheric simulation. The ionisation fraction increases across the shock however the medium remains mostly neutral (Figure \ref{fig:lowchromocontext}d). The intermediate shock that exists within the substructure of the slow-mode shock is visible in Figure \ref{fig:lowchromocontext}b by the slight reversal of the magnetic field across the interface.

The qualitative shock substructure is similar to the mid chromospheric case, with the finite-width of the shock determined entirely by the region where the velocities are decoupled, see Figure \ref{fig:lowchromosub}. The same shock transitions exist, namely a large-scale switch-off slow-mode shock that contains both a sonic shock and a 2-4 intermediate shock within its finite width (Figure \ref{fig:lowchromosub}b). 

Whilst the ionisation rates within the shock increase by several orders of magnitude, the maximum rate within the shock is comparable to the recombination rate outside of the shock, as shown in Figure \ref{fig:lowchromosub}c. As such, the timescales for the ionisation/recombination process remains fairly constant across the domain. This is in contrast to the upper and mid chromospheric cases where the rates were significantly enhanced within the shock, compared to the background values. In all simulations, the temperature increase inside the shock reduces the collisional recombination rate, and increases the ionisation rates. The same behaviour is seen here however the increased values for the ionisation rates remain comparable to the background collisional recombination rate. The lower chromospheric case is highly collisional, limiting the effectiveness of the radiative ionisation.

As with the previous cases, both the post-shock and maximum temperature are significantly lower than the MHD case. This is a result of the increase in ionisation fraction (and level populations of excited neutral levels) across the interface. The losses due to collisions resulting in a macroscopic loss of energy from the plasma that is greater than the recombinative heating. Here the level populations of excited hydrogen become highly significant post-shock (Figure \ref{fig:lowchromosub}d). As such, a significant proportion of energy is lost due to collisional excitation, however this is still much smaller than the energy lost through ionisation from ground state. 


\section{Discussion}

\subsection{Temperature jumps and compression at different heights}

\begin{figure}
    \centering
    \includegraphics[width=0.95\linewidth,clip=true,trim=1.0cm 0.3cm 1.6cm 1.8cm]{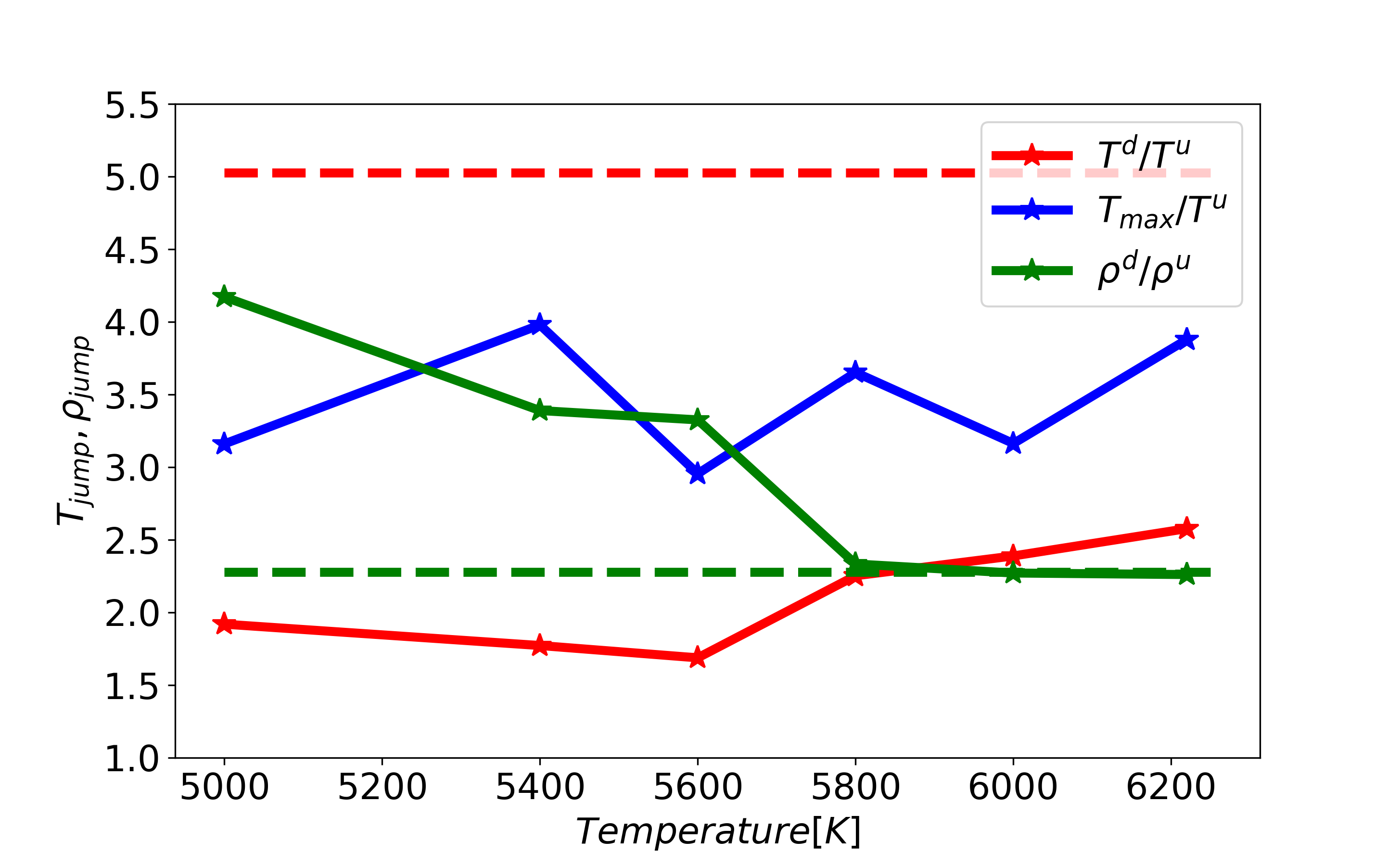}
    \caption{Temperature (red) and density (green) jumps across the shock for different initial reference temperatures. The blue line shows the maximum plasma temperature jump within the finite width of the shock. The red and green dashed lines show the temperature and density jumps in the MHD simulation.}
    \label{fig:tcomp}
\end{figure}

\begin{figure}
    \centering
    \includegraphics[width=0.95\linewidth,clip=true,trim=0.2cm 0.3cm 0.0cm 1.6cm]{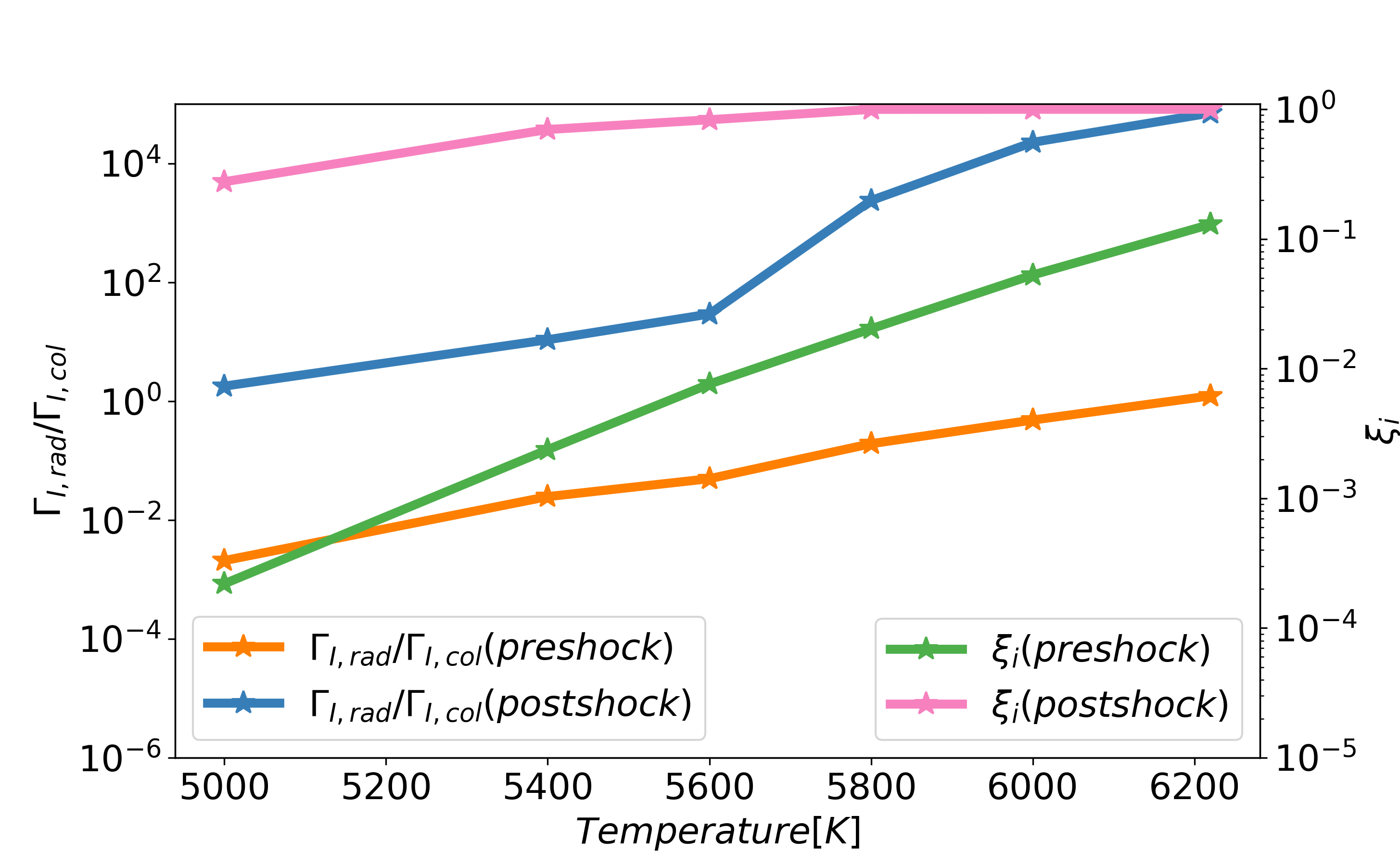}
    \caption{(left axis) Ratio of radiative to collisional ionisation evaluated preshock (orange) and postshock (blue) for different reference temperatures. (right axis) Ionisation fraction evaluated preshock (green) and postshock (pink) for different reference temperatures.}
    \label{fig:tcompion}
\end{figure}

\begin{figure}
    \centering
    \includegraphics[width=0.95\linewidth,clip=true,trim=0.2cm 0.3cm 0.0cm 1.6cm]{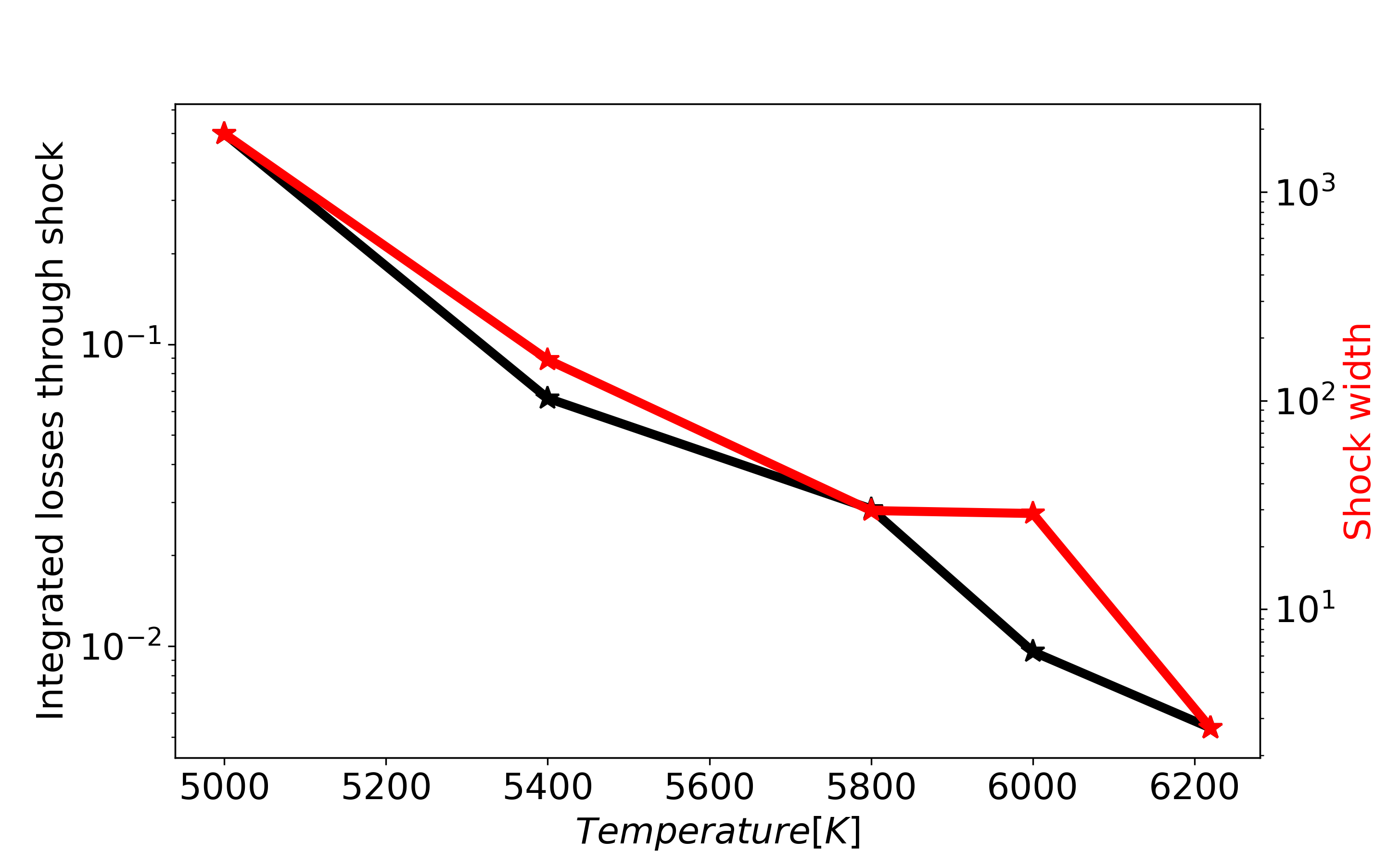}
    \caption{(black line, left axis) Integrated cooling ($\Phi_{I}-\Phi_{R}$) through the shock. (red line, right axis) Finite width of the shock.}
    \label{fig:tcomploss}
\end{figure}

The results presented thus far show that this model has significantly different jumps across the shock than the MHD solution. Of particular interest are the compression ratio and temperature jumps which may have consequences for observations of shocks in the lower solar atmosphere. In this section, a parameter study is performed to study the jump conditions across the interface. In our model, there are two physical parameters that can be investigated: reference temperature and reference electron number density. From the VALCIII model, the electron number density is not expected to vary greatly in the mid to upper chromosphere so this is set to constant as $n_e=7.5\times 10^{16}$ m$^{-3}$ and the reference temperature is varied. It should be emphasised that despite the initial electron number density remaining constant, the total density in each simulation changes depending on the neutral fraction. A set of simulations have been performed with reference temperatures in the range $5000 < T_0 \leq 6200$.

Figure \ref{fig:tcomp} shows the compression ratio and temperature jumps for different reference temperatures, using the same reference electron number density. A consistent feature across all simulations is that the post-shock temperature in the two-fluid simulations is significantly cooler than the MHD result. Furthermore, below $T\approx 5800$ K, the compression across the shock is significantly larger than in the MHD case.

Above $T\approx 5800$ K, the post-shock medium becomes almost entirely ionised with the majority of the post-shock ionisation occurring from the radiative field (Figure \ref{fig:tcompion}), and the large-scale shock jumps are very similar to the MHD result (for example, see Section \ref{sec:upshock}). However, the post-shock temperature in these simulations is significantly lower than the MHD result. In our model, radiative ionisation is an energy-neutral process; the photon provides the exact energy required to release the bound electron such that the internal energy does not change, however the internal energy is now shared across both the electron and proton. As such, the ion (or electron) temperature is half the original neutral temperature. This simplified explanation helps explain why the temperature is less than would be predicted from ideal MHD, however, it is important to note that the simulations presented here have additional physical processes that affect temperature (e.g., the adiabatic compression across the shock).

A particle passing through the shock will experience cooling due to the net loss of energy as a consequence of ionisation. The integrated cooling of such a particle is shown in Figure \ref{fig:tcomploss}, along with the finite width of the shock. Both these quantities reduce exponentially as the temperature increases. For the higher temperatures, radiative ionisation occurs more readily than collisional ionisation, therefore, there is less energy loss, since only collisional ionisation removes electron energy from the macroscopic fluid. This result is in contrast to \cite{Snow2021} where the cooling was inversely proportional to the width of the shock. The reason for this change in trend is that the heating terms are self-consistently modelled here \citep[whereas the heating is assumed constant in][]{Snow2021}, and that a significant proportion of ionisation comes from the radiative field which has no directly associated macroscopic energy loss \citep[radiative ionisation was not considered in][]{Snow2021}.

By their nature, shocks are compressible features that lead to temperature increases due to both adiabatic (compressional) and non-adiabatic (e.g., dissipative) effects that, in the absence of losses, will result in an increase in ionisation fraction. However, the process of collisional ionisation requires energy deposited by free electrons to release bound electrons, resulting in a macroscopic loss of thermal energy from the plasma species. If this process occurs radiatively, then the energy comes from the radiative field and there is no directly associated energy loss from the plasma. Low in the atmosphere, the media is dense and the collisional ionisation rate is much larger than the radiative ionisation, both preshock and postshock, see Figure \ref{fig:tcompion}. As the reference electron temperature increases, the ratio of radiative to collisional ionisation becomes greater. As such, the radiative field is providing a greater proportion of the energy required to ionise the medium, leading to a smaller loss term, and hence less postshock cooling. 
At the same time, as the temperature increases, the initial ionisation fraction increases and hence there is less potential to ionise (i.e., a fully ionised plasma would have no ionisation losses). Note that the energy required for the radiation field comes from the atomic processes, namely spontaneous decay and recombination, and stimulated de-excitation, which all release a photon into the local environment. 

\subsection{Comparison to previous models}

The initial conditions studied in this paper have been extensively studied for different models of coupling between the ion and neutral species. \cite{Hillier2016,Snow2019} studied thermal collisions only (i.e., no ionisation and recombination). In that setup, the jump conditions across the shock reduce to MHD sufficiently far from the shock. As such, the two-fluid effects are localised within the finite-width of the shock. A model that included empirical formulae for collisional ionisation and recombinaton and included non-conservative terms for energy lost during ionisation and an arbitrary heating term was studied in \cite{Snow2021}. Their results showed that the non-conservative terms fundamentally change the permissible shock jumps and a compressible partially-ionised shock must cool across the interface.

The model presented in this paper uses a significantly more-advanced model \citep[namely, the model presented in][adapted for our two-fluid code]{Leenaarts2007,Sollum1999,Johnson1972} for the interactions between ionised and neutral species by using a multi-level hydrogen model that includes both collisional and radiative ionisation and recombination, and self-consistent heating and cooling terms (see Sections \ref{sec:ionmodel}-\ref{sec:radratsec} for full details and assumptions). Similar to \cite{Snow2021}, the non-conservative terms in the plasma energy equation fundamentally change the shock jump conditions and allow compression of the plasma beyond the MHD limit. 

The results of \cite{Snow2021} showed that a partially-ionised shock must cool across the shock interface, however, the simulations presented in this paper all feature a temperature increase across the shock, which is significantly less than the MHD temperature jump. This is a result of the self-consistent heating term being determined by the work done on the electron during recombination and de-excitation. Furthermore, the ionisation/excitation can occur from excited neutral states (requiring less energy than ionisation from ground state) or from the radiative field (no directly associated loss of plasma energy). For the upper chromospheric case, the bulk of the ionisation within the shock is caused by the radiation field.

Whilst the temperature across the shock increased in all simulation presented here, one can theoretically consider cases where the temperature may decrease, in line with the prediction of \cite{Snow2021}. If the radiative recombination is significantly stronger than the collisional recombination, then the heating term will become negligible. Therefore, a system that is dominated by collisional ionisation (which causes a direct energy loss from the plasma) and radiative recombination (which causes no direct heating) then shocks may cool across the interface.

From the analysis in this paper, one would expect that partially-ionised shocks in the lower solar atmosphere behave significantly different to their ideal MHD analogues; the jump conditions do not follow the Rankine-Hugoniot conditions in the lower to mid chromosphere. This is not entirely unexpected since the underlying equations feature cooling and heating terms that can become large, resulting in a departure from ideal MHD results. In particular, we highlight in this paper that the expected temperature jump is significantly cooler and the density jump is much greater than would be predicted using MHD jump conditions.

In this paper, the excess ionisation and recombination energy during the radiative transitions is neglected, with the assumption that radiative ionisation and recombination occurs with the exact energy required for the transition. This neglects the high energy tail which would deposit energy on the free electron during photoionisation leading to an additional heating process, and a cooling process for the photorecombination. This excess energy may become significant when the radiative rates are larger than the collisional rate, e.g., the upper chromosphere. Future work will be to evaluate the consequences of this term.





\subsection{Observational consequences}

Observed shocks usually are classified using the Rankine-Hugoniot shock jump conditions \citep[e.g.,][]{Ruan2018,Anan2019,French2023}. However, the results of this paper show that these jump conditions may not hold in the lower solar atmosphere where the medium is partially ionised. The Rankine-Hugoniot conditions are derived from the ideal MHD equations, however, these underlying equations become less suitable when non-ideal source terms become large. In the lower solar atmosphere, the energy required to ionise the medium represents a significant loss term in the energy equation and fundamentally alters the permissible shock jumps. As such, applying the Rankine-Hugoniot shock jump conditions to shocks observed in the lower solar atmosphere may not be appropriate and may result in incorrect classifications. 


The expected emission from lower atmospheric shocks may be different to the MHD predicted values due to the much larger increase in density and lower temperature jumps in the partially ionised model. 
The extent of the observed non-thermal line broadening in shocks is not fully explained by the non-equilibrium ionisation \citep{Pontieu2015}. The temperature reduction and enhanced compression shown to exist within the two-fluid shock studied here may provide a potential explanation. 

All simulations show a large increase in excited states of neutral hydrogen across the interface. The greatest increase is in the first excited state, which would manifest observationally as an increase in Lyman emission at the shock front. Lyman emission has been comprehensively studied in the context of flares \citep{Druett2019,Mclaughlin2023} and a similar spike in emission may be observable in chromospheric shocks. 
Whilst the high optical depths of the Lyman line wavelengths would make this emission difficult to directly observe in the chromosphere, the increase of neutral hydrogen with electrons in higher excited states may lead to observable increases in opacity and emission in the Balmer (or other hydrogen) lines.




All conditions studied in this paper result in significantly less post-shock heating than the MHD equivalents. This can be compared and contrasted to the radiative simulations of \cite{Martinez2020}, where the heating through ambipolar diffusion was studied in 2.5D simulations. They found that shocks were heated to greater temperatures, with an adiabatic post-shock expansion producing cool post-shock bubbles. In this paper, we see that both the shock substructure and post-shock regions are cooler than their MHD analogues due to non-adiabatic cooling as a result of the thermal energy lost during ionisation. There are significant differences between the underlying models used which limits direct comparison \citep[most notably, ][have a stratified atmosphere and include helium in their model]{Martinez2020}. The main advantage of the model we have presented is that the two-fluid framework allows for study of the shock substructure and reveals that cooling occurs both postshock, and within the finite width of the shock. The maximum temperature jump obtained within the shock is cooler than the MHD temperature. 


\section{Conclusions}

In this paper we have studied reconnection-like shocks in the lower solar atmosphere through two-fluid numerical simulations with collisional and radiative ionisation and recombination. The underlying mode is built around a multi-level hydrogen model presented by \citep{Sollum1999}. Three atmospheric heights are studied in detail spanning the upper, middle and lower chromosphere. In all cases, the two-fluid simulations show significantly enhanced compression across the shock and a lower post-shock temperature compared to the MHD result. 

The main result is that two-fluid partially-ionised shocks behave very differently to their MHD analogues. As such, the Rankine-Hugoniot shock jump conditions that are often used to classify observed shocks may not be suitable for shocks in the solar chromosphere. This has implications for inferred dissipation of shocks \citep[e.g.,][]{Anan2019} and the role of shocks in heating the lower solar atmosphere.

The collisionallity of the medium strongly affects both the post-shock and sub-shock regions. Lower in the atmosphere, the medium is strongly collisional and thus the radiative rates are less important and the losses due to ionisation and excitation are strong, leading to a reduced temperature. In the upper chromosphere, the radiative ioniation provides significant energy for the ionisation process within the shock and hence the cooling is less efficient than in lower atmospheric heights. 

Non-equilibrium ionisation effects play a fundamental role in the temperature evolution of partially-ionised shocks. The model presented in this paper shows that partially-ionisation results in a cooler temperature in the post-shock region (and within the shock substructure) than would be predicted from MHD models. As such, two-fluid shocks must be studied to understand the role of shocks in chromospheric heating. 

\section*{Acknowledgements}

BS and AH are supported by STFC research grant ST/R000891/1 and ST/V000659/1. MD is supported by FWO project G0B4521N. 
We would like to thank Rob Rutten, Jorit Leenaarts and Giulio Del Zanna for stimulating discussions regarding ionisation and recombination. We are saddened to hear of Rob Rutten's passing. We would like to thank the keen insight of the referee. 
The authors would like to acknowledge the use of the University of Exeter High-Performance Computing (HPC) facility in carrying out this work.
\section*{Data Availability}

The simulation data from this study are available from BS upon reasonable request. The (P\underline{I}P) code is available at \href{https://github.com/AstroSnow/PIP}{https://github.com/AstroSnow/PIP}.



\bibliographystyle{mnras}
\bibliography{bib} 




\appendix

\section{Collisional ionisation and excitation coefficients}

\begin{table}
    \centering
    \caption{Oscillator strengths From Table 1 in \protect\cite{Goldwire1968}}
    \begin{tabular}{c|c}
        Level transition (lower,upper) & Oscillator strength \\
        1,2 & 4.1619672e-1 \\
        1,3 & 7.9101563e-2 \\
        1,4 & 2.8991029e-2 \\
        1,5 & 1.3938344e-2 \\
        2,3 & 6.4074704e-1 \\
        2,4 & 1.1932114e-1 \\
        2,5 & 4.4670295e-2 \\
        3,4 & 8.4209639e-1 \\
        3,5 & 1.5058408e-1 \\
        4,5 & 1.0377363e0
    \end{tabular}
    \label{tab:oscst}
\end{table}

\begin{table}
    \centering
    \caption{Constants used in the calculations}
    \begin{tabular}{p{2cm} | c c }
        Constant & Symbol & Value [units] \\
        \hline 
        Ion-neutral collisional cross-section & $\Sigma_{in}$ & $5\times10^{-19}$ [m$^2$]
        \\ Hydrogen mass & $M$ & $1.6735575 \times 10^{-27}$ [kg]
        \\ Electron mass & $m_e$ & $9.10938356 \times 10^{-31}$ [kg]
        \\ Planck's constant & $h$  & $6.62607004 \times 10 ^{-34}$ [m$^2$ kg s$^{-1}$]
        \\ Speed of light & $c$ & $299792458$ [m s$^{-1}$]
        \\ Bohr radius & $a_0$ & $5.29 \times 10^{-11}$ [m]
        \\ Electron charge & $e$ & $-1.6 \times 10^{-19}$ [Coulombs] 
        \\ Boltzmann constant & $k_B$ & $1.38064852 \times 10^{-23}$ [m$^2$kg s$^{-2}K^{-1}$]
    \end{tabular}
    \label{tab:constants}
\end{table}


\subsection{Rate coefficients} \label{app:ratecoef}

It should be emphasised that here the complete set of equations for our rate coefficients is listed for completeness. For physical insight into these equations, we would advise the reader to analyse the source texts \citep[e.g.,][]{Mihalis1978,Druett2018}.
Rate coefficients $C_{\rm{exc}},C_{\rm{ion}}$ that are used in this model come from \cite{Johnson1972}.

\subsubsection{Excitation coefficient}

For an excitation from level $i$ to level $j$, the excitation coefficients can be calculated according to \cite{Johnson1972} as
\begin{gather}
    C_{\rm{exc}}= \sqrt{\frac{8 k_B T_e}{\pi m_e}} \frac{2 \pi a_0^2 \hat{y}^2 i^2}{x_{r}} \left[ A_{i,j} \left( \left( \frac{1}{\hat{y}} +\frac{1}{2} \right) E_{1}(\hat{y}) - \left( \frac{1}{\hat{z}}+\frac{1}{2}\right)E_{1}(\hat{z})\right) \right. \nonumber \\ \hspace{2cm}\left. \left(B_{i,j}-A_{i,j} \log{\frac{2 i^2}{x_r}}\right) \left(\frac{E_{2}(\hat{z})}{\hat{y}}-\frac{E_{2}(\hat{z})}{\hat{z}}\right) \right] \label{eqn:cexc}
\end{gather}
for Boltzmann constant $k_B=1.38064852 \times 10^{-23}$ m$^2$ kg s$^{-1}$, electron mass $m_e=9.10938356 \times 10^{-31}$ kg, and Bohr's radius $a_0=5.29\times 10^{-11}$ m. The input variables are the electron number density $n_e$ and the electron temperature $T_e$.  Further, we define:
\begin{gather}
    \hat{y}=E_{i,j}/k_B T_e \\
    \hat{z}=r_{i,j} +E_{i,j}/k_B T_e \\
    r_{i,j}=r_i x_r \\
    x_r=1-\frac{i^2}{j^2}
\end{gather}
where $E_{i,j}$ is the change in ionisation energy from level $i$ to level $j$. $r_i$ is calculated according to:
\begin{gather}
    r_1=0.45 \\
    r_i=1.94 i^{-1.57} \mbox{for $i>1$}
\end{gather}
The exponential integrals are defined as
\begin{gather}
    E_n(x)=\int ^\infty _1 e^{-x\omega} \omega^{-n} d\omega
\end{gather}
A pre-calculated table is used in the numerical simulations for computational speed.

\subsection{Ionisation coefficient}

Similarly, we use the ionisation coefficients provided by \cite{Johnson1972} to calculate our transitions from neutral level $i$ to charged state $c$
\begin{gather}
    C_{\rm{ion}}= \sqrt{\frac{8 k_B T_e}{\pi m_e}} 2 \pi a_0^2 y_n^2 i^2 \left[ A_{n0}\left(\frac{E_1(y)}{y_n}-\frac{E_1(z)}{z_n} \right) + \right. \nonumber \\ \hspace{2cm}\left.  \left( b_{n0}-A_{n0} \log(2i^2) \right) \left(\zeta_{y}-\zeta_{z} \right) \right]
\end{gather}
where
\begin{gather}
    y_n=\frac{\phi_i}{k_B T_e} \\
    z_n=r_c + \frac{\phi_i}{k_B T_e} \\
    \zeta_y=E_0{y_n}-2E_1(y_n)+E_2(y_n) \\
    \zeta_z=E_0{z_n}-2E_1(z_n)+E_2(z_n) \\
    A_{n0}= \frac{32 i}{3 \sqrt{3} \pi} + \frac{g_1}{3}+ \frac{g_2}{4} + \frac{g_3}{5} \\
    B_{n0}= \frac{2i^2}{3} \left( 5+ b_n \right) \\
    b_n= 
\begin{cases}
    -0.603,& \text{if } i = 1\\
    \frac{1}{i} \left(4-\frac{18.63}{i}+\frac{36.24}{i^2}- \frac{28.09}{i^3} \right),  & \text{otherwise}
\end{cases}
\end{gather}
The factors $g_1,g_2,g_3$ are defined from \cite[][Table 1]{Johnson1972} as
\begin{gather}
        g_1= 
\begin{cases}
    1.330,& \text{if } i= 1\\
    1.0785,& \text{if } i= 2\\
    0.9935+\frac{0.2328}{i}-\frac{0.1296}{i^2}, & \text{otherwise}
\end{cases} \\
        g_2= 
\begin{cases}
    -0.4059,& \text{if } i= 1\\
    -0.2319,& \text{if } i= 2\\
    \frac{-1}{i} \left( 0.6282 -\frac{0.5598}{i} + \frac{0.5299}{i^2} \right) & \text{otherwise}
\end{cases}\\
        g_3= 
\begin{cases}
    0.07014,& \text{if } i= 1\\
    0.02947,& \text{if } i= 2\\
    \frac{1}{i^2} \left( 0.3887 -\frac{1.181}{i} + \frac{1.470}{i^2} \right) & \text{otherwise}
\end{cases}
\end{gather}

\section{Frictional heating and thermal damping terms} \label{app:heatdamp}

\subsection{Thermal damping}

Thermal damping acts to thermally equalise the system such that the plasma and neutrals have the same temperature at infinite times. This equalisation occurs collisionally in the form
\begin{gather}
    \alpha _c \rho_{\rm n} \rho_{\rm p} \frac{1}{\gamma(\gamma -1)} (T_{\rm n}-T_{\rm p})
\end{gather}
A similar effect occurs through ionisation and recombination of the form
\begin{gather}
    \frac{1}{\gamma(\gamma -1)} (\Gamma_I \rho_{\rm n} T_{\rm n}- \Gamma_R \rho_{\rm p} T_{\rm p})
\end{gather}
Both these equations are evident in the energy equations. 

\subsection{Frictional heating}

A heating term occurs due to friction which can be calculated as the difference between the energy term and the work done. The work done is calculated as the dot product of the velocity with the momentum equation. The work done by neutral collisions is
\begin{gather}
    \textbf{v}_{\rm n} \cdot (\alpha \rho_{\rm n} \rho_{\rm p} (\textbf{v}_{\rm n}-\textbf{v}_{\rm p})) = \alpha \rho_{\rm p} \rho_{\rm n} (\textbf{v}_{\rm n}^2 - \textbf{v}_{\rm p} \cdot \textbf{v}_{\rm n})
\end{gather}
The frictional heating is then calculated as the difference between energy term and work done
\begin{gather}
    H_f=\frac{1}{2}\alpha \rho_{\rm n}\rho_{\rm p} \left(v_{\rm n}^2-v_{\rm p}^2\right)-\alpha \rho_{\rm n} \rho_{\rm p} \left( \textbf{v}_{\rm n}^2-\textbf{v}_{\rm n} \cdot \textbf{v}_{\rm p} \right) \\
    =\frac{1}{2} \alpha \rho_{\rm n} \rho_{\rm p} \left( \textbf{v}_{\rm n}-\textbf{v}_{\rm p}\right)^2
\end{gather}
This term is equal for both the neutrals and ions.

Similar heating terms exist due to ionisation and recombination. The work done through ionisation and recombination is
\begin{gather}
    \textbf{G}=\Gamma_R \rho_{\rm p} \textbf{v}_{\rm p} - \Gamma_I \rho_{\rm n} \textbf{v}_{\rm n} \\
    \textbf{v}_{\rm n}\cdot \textbf{G}= \Gamma_R \rho_{\rm p} \textbf{v}_{\rm p} \cdot \textbf{v}_{\rm n} - \Gamma_I \rho_{\rm n} \textbf{v}_{\rm n}^2 \\
    \textbf{v}_{\rm p}\cdot -(\textbf{G})= \Gamma_I \rho_{\rm n} \textbf{v}_{\rm p} \cdot \textbf{v}_{\rm n}-\Gamma_R \rho_{\rm p} \textbf{v}_{\rm p}^2 
\end{gather}
which leads to the heating terms
\begin{gather}
    H_{\rm n}=\frac{1}{2} \left(\Gamma_R \rho_{\rm p} \textbf{v}_{\rm p}^2 - \Gamma _I \rho_{\rm n} \textbf{v}_{\rm n}^2\right)-\left( \Gamma_R \rho_{\rm p}\textbf{v}_{\rm p}\cdot \textbf{v}_{\rm n}-\Gamma_I \rho_{\rm n} \textbf{v}_{\rm n}^2 \right) \\
    \hspace{1cm}=\frac{1}{2}\left( \Gamma_R\rho_{\rm p}\textbf{v}_{\rm p}^2 -2\Gamma_R \rho_{\rm p} \textbf{v}_{\rm p} \cdot \textbf{v}_{\rm n} + \Gamma_I \rho_{\rm n} \textbf{v}_{\rm n}^2 \right)\\
    H_{\rm p}=\frac{1}{2} \left(\Gamma_I \rho_{\rm n} \textbf{v}_{\rm n}^2 - \Gamma _R \rho_{\rm p} \textbf{v}_{\rm p}^2\right)-\left(\Gamma_I \rho_{\rm n} \textbf{v}_{\rm p}\cdot \textbf{v}_{\rm n} -\Gamma_R \rho_{\rm p} \textbf{v}_{\rm p}^2 \right) \\
    \hspace{1cm}=\frac{1}{2}\left( \Gamma_R\rho_{\rm p}\textbf{v}_{\rm p}^2 -2\Gamma_I \rho_{\rm n} \textbf{v}_{\rm p} \cdot \textbf{v}_{\rm n} + \Gamma_I \rho_{\rm n} \textbf{v}_{\rm n}^2 \right)
\end{gather}
These terms have several interesting features. The heating is slightly different in the ion and neutral fluids, unlike the analogous term due to thermal collisions. In the limit of ionisation/recombination equilibrium (that is $\Gamma_I \rho_{\rm n} = \Gamma_R \rho _{\rm p} $), then these heating terms become equal and can be expressed in the form $H_I=H_R=\frac{1}{2} \Gamma_I \rho_{\rm n} (v_{\rm p}-v_{\rm n})^2$ which is always positive. Outside of local ionisation/recombination equilibrium, the heating terms can become negative, for example when the ionisation rate is negligibly small and the neutral velocity is large. 

Adding these terms together, one yields the total frictional heating due to ionisation and recombination
\begin{gather}
    H_{\rm n}+H_{\rm p}= \Gamma_R \rho_{\rm p} \textbf{v}_{\rm p}^2 -\left( \Gamma_R \rho_{\rm p}+\Gamma_I \rho_{\rm n} \right) \textbf{v}_{\rm n}\cdot \textbf{v}_{\rm p} +\Gamma_I \rho_{\rm n} \textbf{v}_{\rm n}^2
\end{gather}

\bsp	
\label{lastpage}
\end{document}